\documentclass[%
twocolumn,
superscriptaddress,
 amsmath,amssymb,
 aps,
 prb,
]{revtex4-2}

\usepackage{graphicx}
\usepackage{dcolumn}
\usepackage{bm}
\usepackage{color}
\usepackage{xspace}
\usepackage{multirow}
\usepackage{ulem}
\usepackage{hyperref}
\hypersetup{
    colorlinks=true,
    linkcolor=blue,
    urlcolor=blue,
    citecolor=blue
}
\usepackage{here}

\newcommand{\red}[1]{\textcolor{red}{#1}}
\newcommand{\blue}[1]{\textcolor{blue}{#1}}

\definecolor{green}{rgb}{0,0.6,0.1}
\newcommand{\green}[1]{\textcolor{green}{#1}}

\renewcommand{\red}[1]{#1}
\renewcommand{\blue}[1]{#1}

\renewcommand{\green}[1]{#1}

\newcommand{\rr}[1]{\textcolor{black}{#1}}

\newcommand{\rrr}[1]{\textcolor{black}{#1}}

\newcommand{\NNO}{${\rm NdNiO}_{2}$\xspace}
\newcommand{\xx}{$d_{x^2-y^2}$\xspace}
\newcommand{\xy}{$d_{xy}$\xspace}

\newcommand{\QE}{{\textsc{Quantum ESPRESSO}}\xspace}
\newcommand{\pti}{$AB_{2}$NiO$_{3}$\xspace}
\newcommand{\ptii}{$AB_{2}$NiO$_{4}$\xspace}

\DeclareMathAlphabet{\mathbit}{OT1}{cmr}{bx}{it}

\begin{document}

\preprint{APS/123-QED}

\title{Materials design of dynamically stable $d^9$ layered nickelates}

\author{Motoaki Hirayama}
\email{motoaki.hirayama@riken.jp}
\affiliation{ 
RIKEN Center for Emergent Matter Science, 2-1 Hirosawa, Wako, Saitama 351-0198, Japan 
}


\author{Terumasa Tadano}
\email{TADANO.Terumasa@nims.go.jp}
\affiliation{
Research Center for Magnetic and Spintronic Materials, National Institute for Materials Science, Tsukuba 305-0047, Japan
}

\author{Yusuke Nomura}
\affiliation{ 
RIKEN Center for Emergent Matter Science, 2-1 Hirosawa, Wako, Saitama 351-0198, Japan 
}

\author{Ryotaro Arita}
\affiliation{ 
RIKEN Center for Emergent Matter Science, 2-1 Hirosawa, Wako, Saitama 351-0198, Japan 
}
\affiliation{
Department of Applied Physics, The University of Tokyo,7-3-1 Hongo, Bunkyo-ku, Tokyo 113-8656
}


\date{\today}

\begin{abstract}
Motivated by the recent discovery of superconductivity in the Sr-doped layered nickelate \NNO, we perform a systematic computational materials design of layered nickelates that are dynamically stable and whose electronic structure better mimics the electronic structure of high-$T_c$ cuprates than \NNO. 
While the Ni $3d$ orbitals are self-doped from the $d^9$ configuration in \NNO 
and the Nd-layer states form Fermi pockets,
we find more than 10 promising compounds for which the self-doping is almost or even completely suppressed.
We derive effective single-band models for those materials and find that they are in the strongly-correlated regime. We also investigate the possibility of palladate analogs of high-$T_c$ cuprates. Once synthesized, these nickelates and palladates will provide a firm ground for studying superconductivity in the Mott-Hubbard regime of the Zaanen-Sawatzky-Allen classification.
\end{abstract}


\maketitle


\section{Introduction} 
The recent discovery of superconductivity in the Sr-doped nickelate NdNiO$_2$~\cite{Li_2019}
offers a new exciting platform for investigating unconventional superconductivity in layered correlated materials, thus prompting active theoretical and experimental studies~\cite{Botana_arXiv,Sakakibara_arXiv,Hirsch_2019,Yamagami_2019, Hepting_2019,Mi_arXiv,Wu_arXiv,Nomura_arXiv,Gao_2019,Ryee_2019,HuZhang_arXiv,Singh_2019,Zhang_2019selfdoped,Zhang_arXiv, Jiang_2019,Werner_2019,Hu_arXiv}.
Here, one fascinating avenue to explore is computational materials design of new nickelate superconductors exploiting the layered crystal structure.
Since \NNO has alternate stacks of the superconducting NiO$_2$ layer and the Nd layer, we can design different nickelate superconductors by replacing the Nd layer. Indeed, for layered superconductors such as high transition-temperature ($T_c$) cuprates~\cite{Bednorz_1986}, iron-based superconductors~\cite{kamihara_2008}, BiS$_2$ superconductors~\cite{Mizuguchi_2012}, a rich variety of families exhibiting different physical properties have been discovered. While superconductivity has been observed only in the Sr-doped NdNiO$_2$ so far, discovery of other nickelate superconductors would have a great impact on the studies not only of nickelates but also of other layered transition-metal oxides. 

Although the low-energy electronic structure of \NNO is quite similar to that of high-$T_c$ cuprates 
\rr{in the sense that only the Ni $3d_{x^2-y^2}$ orbital among the five $3d$ orbitals forms the Fermi pocket,}
there is a distinct difference; extra states cross the Fermi level and induces the self-doping of the Ni 3\xx orbital,
\rr{i.e., the occupation of the Ni $3d$ orbitals deviates from $d^9$}
~\cite{Lee_2004,Botana_arXiv,Nomura_arXiv,Zhang_2019selfdoped}.
In particular, the Fermi pocket around the A point in the Brillouin zone is noticeable, which is formed by the bonding orbital between the interstitial $s$ state formed in the Nd layer and the Nd 5\xy state (see Fig.~\ref{NdNiO2_band}).
We have recently shown that the size of the extra Fermi pocket can be reduced by changing the Nd layer with other anion layers~\cite{Nomura_arXiv}. 
This finding motivated us to seek even better compositions and structures that
\rr{realizes a two-dimensional single-band correlated system formed by the Ni $3d_{x^2-y^2}$ electrons}
\footnote{\rr{It should be noted that we can consider other scenarios for the superconductivity in Sr$_{0.2}$Nd$_{0.8}$NiO$_2$, where Nd $d$ electrons play a crucial role (see e.g., Ref.~\onlinecite{Sawatzky19}).}}.

\rr{In fact, designing a two-dimensional single-component correlated electron system has been a key issue to realize an analog of high-$T_c$ cuprates
\footnote{
\rr{For a recent review on designing analogs of cuprates, see, e.g., Ref.~\onlinecite{Norman_2016}.}
}.
While intensive studies for $d^7$ nickelate heterostructures have been performed, it is not an easy task to eliminate the $3d_{3z^2-r^2}$ band from the Fermi level~\cite{Chakhalian14,Middey16,Catalano_2018}.
Although some iridates are promising candidate for $5d$ analog of cuprates~\cite{Kim_2015,Yan_2015},
it is difficult to isolate
the $j_{\text{eff}}=1/2$ bands from the $j_{\text{eff}}=3/2$ bands~\cite{Arita_2012,Martins_2011}.
}

\rrr{In the recently reported superconducting nickelates, if the self-doping can be suppressed, Ni $3d_{x^2-y^2}$ single-band system is realized.
If we follow the ``orbital distillation'' scenario}~
\footnote{For cuprates, the effect of ``orbital distillation'' has been studied and
shown that the isolation of the $3d_{x^2-y^2}$ band from the other bands is important to achieve high $T_c$~\cite{Sakakibara_2010,Sakakibara_2012,Sakakibara_2012b}
},
\rrr{this situation will serve as a $d^9$ (instead of intensively studied $d^7$) analog of the cuprates.}
In this paper, we present a systematic computational search for promising layered nickelates that are dynamically stable and whose electronic structure
\rr{is well described by a single component model.}
When designing possible structures of layered nickelates in a controlled and systematic way, we refer to the seminal work by Tokura and Arima in 1990 for high-$T_c$ cuprates~\cite{Tokura_1990}. There, they introduced the concept of ``block layer'' (BL), 
which is an atomic layer composed of metallic ions and oxygens or halogens, for classifying and designing the crystal structures of high-$T_c$ cuprates.
Among the 14 BLs they proposed, 10 BLs contain apical oxygen, which is harmful to realizing 
the valence of Ni$^{1+}$ (or equivalently $d^9$ configuration).
Therefore, we focus on the other four BLs, which are free from apical oxygens, as a building block of nickelates.
The constituent elements of the four BLs are also generated systematically by considering the requirements of the valence of a BL.

Based on our design principle, we systematically generate 57 new layered nickelates. To make our computational prediction more reliable, we assess the dynamical stability of these nickelates by performing comprehensive phonon calculations and show that 16 nickelates are indeed dynamically stable. For the selected dynamically stable systems, we further derive effective single-band Hamiltonians and compare them with that for \NNO.
We show that all of the theoretically proposed nickelates reside in the strongly correlated regime without forming an extra Fermi pocket as large as that of \NNO. 
Indeed, some nickelates have no extra Fermi pocket other than the Ni 3\xx band and better mimic the band structure of high-$T_c$ cuprates. 
In such nickelates, a single-band Hubbard model will give a good description because \green{the Ni 3\xx band} is well isolated from the other $3d$ bands, and the charge transfer energy is large.
Therefore, they provide, once synthesized, a better platform for studying superconductivity in the Mott-Hubbard regime, 
which will also help the understanding of the superconductivity in high-$T_c$ cuprates realized in the charge-transfer regime.

The structure of the paper is as follows. After briefly discussing the origin of the extra Fermi pocket observed in \NNO in Sec.~\ref{sec_112}, we describe our detailed strategy for the materials design in Secs.~\ref{sec_guide} and \ref{sec_structure}. In Sec.~\ref{sec_phonon}, we identify the \green{dynamically} stable materials by performing \textit{ab initio} phonon calculations. We then present the electronic structure of the stable materials in Sec.~\ref{sec_band} and derive effective single-band models in Sec.~\ref{sec_effective}. In Sec.~\ref{sec:discussion}, we further extend our systematic search to Pd oxides (palladates). Sec.~\ref{sec_conclusion} is devoted to the summary of the present study.

\section{Origin of extra Fermi pocket in N\lowercase{d}N\lowercase{i}O$_{2}$}

\label{sec_112}

\begin{figure}[tbhp]
\vspace{0cm}
\begin{center}
\includegraphics[width=0.4\textwidth]{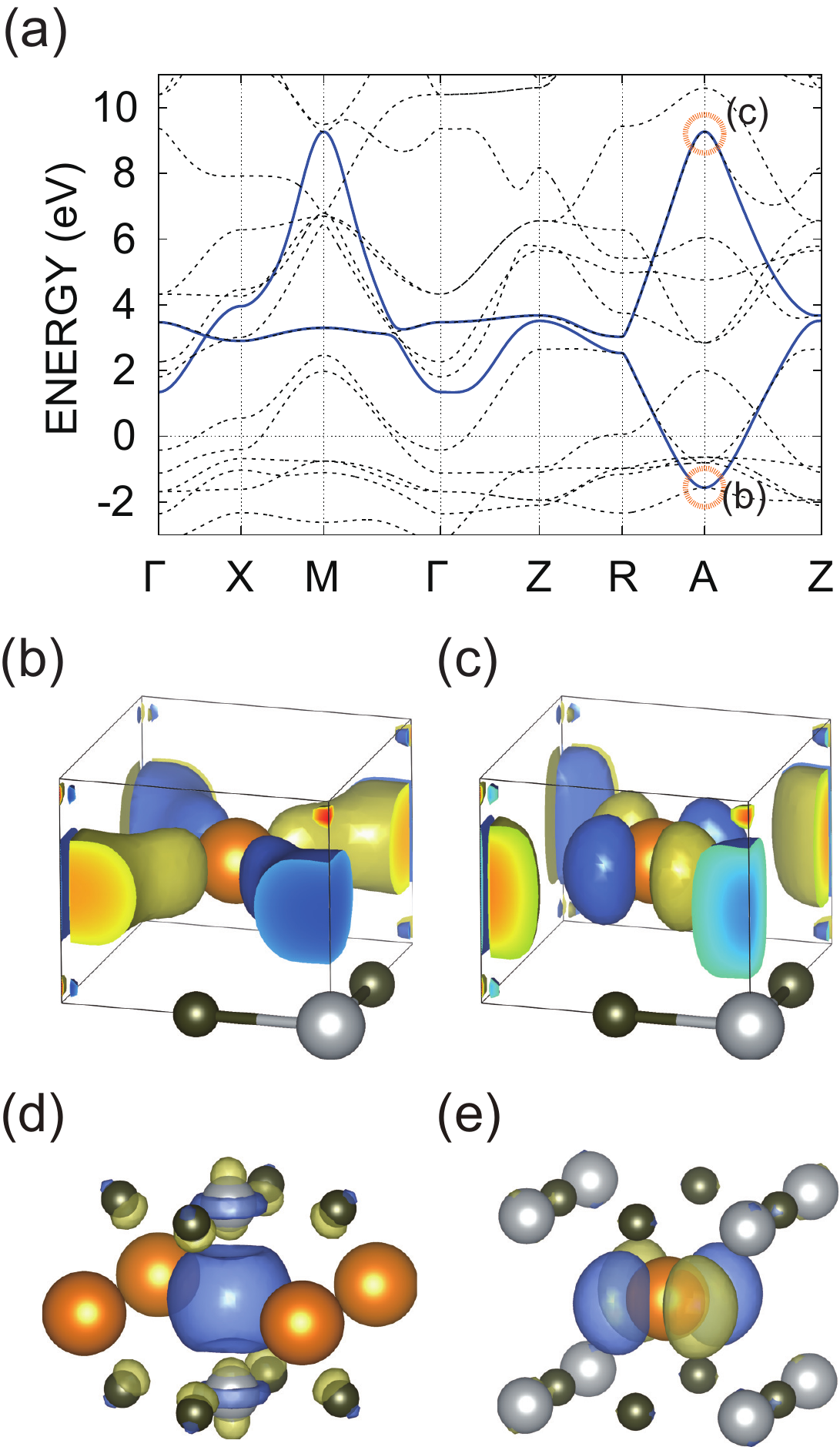}
\caption{
(a) Electronic band structure of NdNiO$_2$ (black dotted line)
and that of a two-orbital tight-binding model for the interstitial $s$ and Nd $5d_{xy}$ Wannier orbitals
(blue solid line).
The energy is measured from the Fermi level.
(b),(c) Bloch function of the band bottom of the bonding band between the interstitial $s$ orbital and the Nd 5$d_{xy}$ orbital and the band top of the antibonding band at the A point, respectively.
The results shown in (b) and (c) correspond to the states indicated by the orange open circles in (a).
(d),(e) Wannier function of the interstitial $s$ orbital and \red{the Nd $5d_{xy}$} orbital, respectively.
The calculation is performed by using OpenMX~\cite{Ozaki_2003} (see Appendix \ref{sec_methods} for the computational details).
}
\label{NdNiO2_band}
\end{center}
\end{figure}

We first discuss the origin of the large extra Fermi pocket in the low-energy electronic structure of \NNO. Figure ~\ref{NdNiO2_band}(a) shows the calculated electronic band structure of \NNO. 
While the electronic structure of \NNO looks very similar to that of high-$T_c$ cuprates in that the Ni 3\xx orbital makes a large two-dimensional Fermi surface, one notable difference exists: the presence of extra states intersecting the Fermi level. 
In particular, the extra Fermi pocket around the A point of the Brillouin zone is noteworthy, which is formed by the bonding state between the interstitial $s$-like orbital and \red{the Nd $5d_{xy}$ orbital} shown respectively in Figs.~\ref{NdNiO2_band}(d) and \ref{NdNiO2_band}(e)
\rr{(The interstitial $s$-like state at the R point is shown in Appendix \ref{sec:interstital}).
}
The hybridization between the bonding state and the Ni 3\xx orbital is small~\cite{Nomura_arXiv}, but the formation of the extra Fermi pocket induce 
the self-doping of the Ni 3\xx orbital~\cite{Lee_2004,Botana_arXiv,Nomura_arXiv, Zhang_2019selfdoped}.

The presence of the extra Fermi pocket can mainly be attributed to the relatively high energy level of the Ni 3\xx orbital of the NiO$_{2}$ layer compared with that of the Cu 3\xx orbital in a CuO$_{2}$ layer. To realize the $d^9$ electron configuration in a NiO$_{2}$ or CuO$_{2}$ plane, the valence of Ni and Cu must be $+1$ and $+2$, respectively. Since the most common oxidation state of these transition metals is $+2$, (NiO$_{2}$)$^{3-}$ is energetically less stable than (CuO$_{2}$)$^{2-}$, making the energy level of \red{the Ni 3\xx orbital higher than that of the Cu 3\xx orbital}.

Since it is difficult to control the energy level of the Ni 3\xx orbital,
our strategy here is to lift the energy level of the bonding orbital around the A point and thereby remove the extra Fermi pocket. 
The energy level of the bonding orbital will be lifted by
\begin{itemize}
    \item raising the energy level of the interstitial $s$ and the cation \xy orbitals.
    \item reducing the band splitting between the bonding and antibonding orbitals.
\end{itemize}
It can be achieved by replacing the Nd layer with other layers, which affects both of the above two factors~\cite{Nomura_arXiv}.

\rr{
  Since the presence of the extra Fermi pockets may result from the generalized-gradient approximation (GGA) employed in the present density functional theory (DFT) calculations, we also investigate their robustness by performing a band structure calculation within the GW approximation (GWA). As shown in Fig.~\ref{LaNiO2_GW_band} of Appendix \ref{sec:GW}, the extra Fermi pocket at the $\Gamma$ point, which originates from the $5d_{3z^2-r^2}$ orbital and exists in the GGA result, disappears within the GWA. By contrast, the size of the Fermi pocket around the A point is almost the same between the GGA and GWA results, thus validating our exclusive focus on the Fermi pocket around the A point in this paper.
}


\section{Choice of block layers}
\label{sec_guide}

\begin{figure}[thbp]
\vspace{0cm}
\begin{center}
\includegraphics[width=0.35\textwidth]{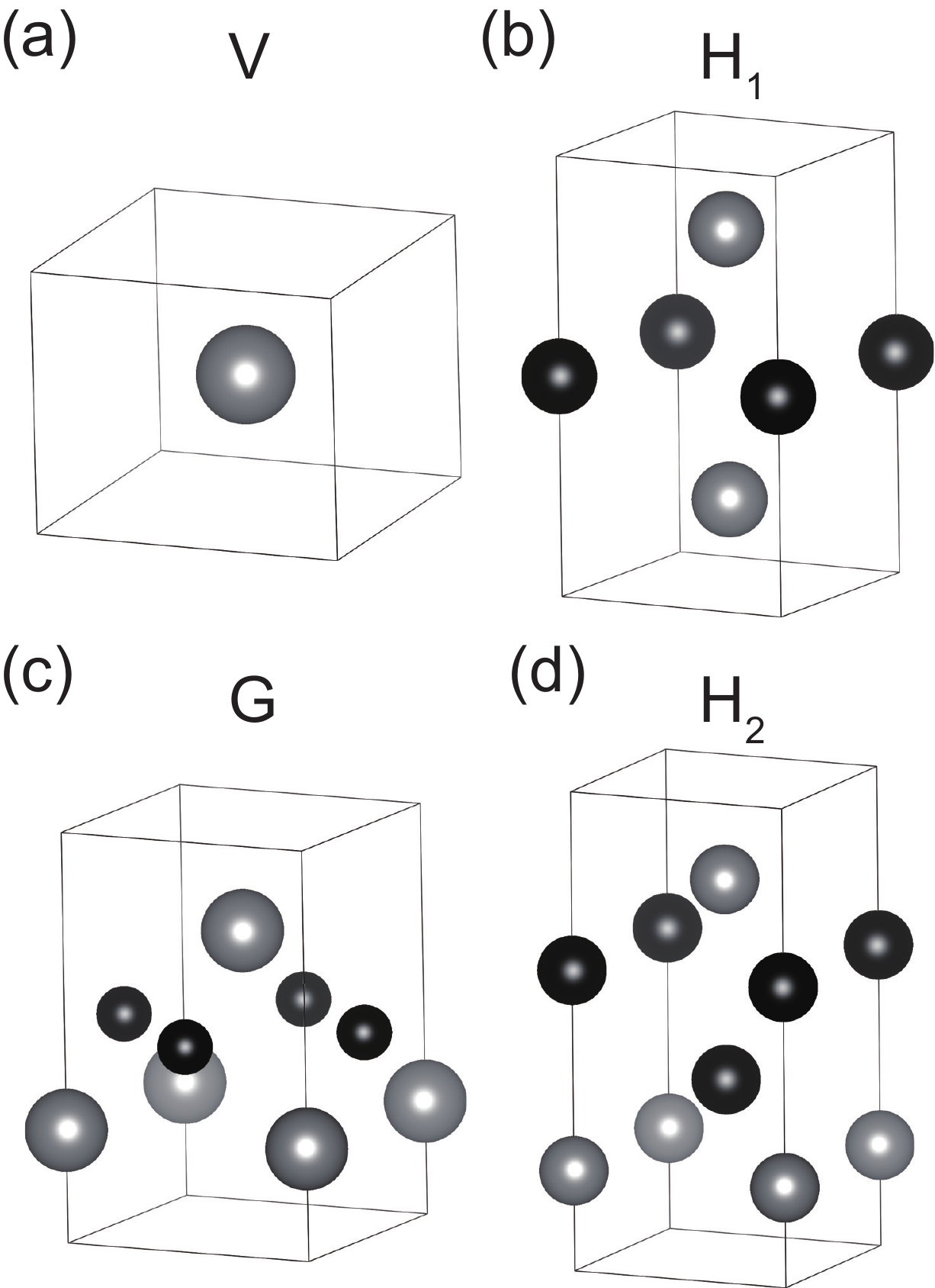}
\caption{
Crystal structure of the block layer:
(a) V-block, (b) H$_1$-block, (c) G-block, and (d) H$_2$-block.
}
\label{block_layer}
\end{center}
\end{figure}

\begin{figure*}[htbp]
\vspace{0cm}
\begin{center}
\includegraphics[width=0.8\textwidth]{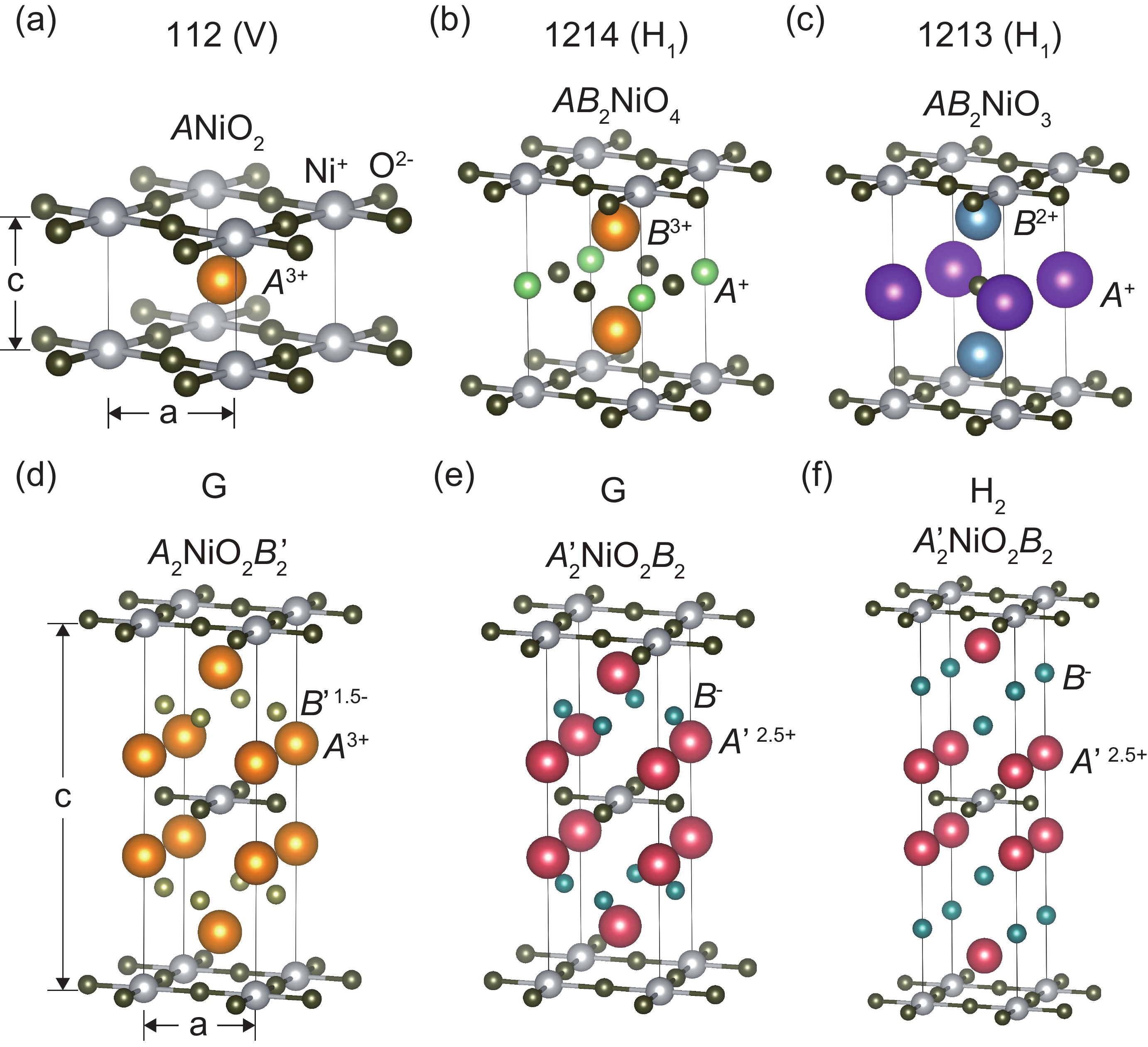}
\caption{
Crystal structures of layered Ni $d^9$ oxides studied in the present study:
(a)$A$NiO$_2$,
(b)$AB_2$NiO$_4$,
(c)$AB_2$NiO$_3$,
(d)$A_2$NiO$_2B_2'$,
(e)$A'_2$NiO$_2B_2$ (G block),
and (f)$A'_{2}$NiO$_2B_2$ (H$_2$ block).
For $A_2$NiO$_2B_2'$, a
half of $B'$ sites are occupied by O$^{2-}$ and the other half are occupied by F$^{-}$ or H$^{-}$ (e.g. O$_{0.5}$F$_{0.5}$).
Another possibility is 
$3/4$ of $B'$ sites are occupied by O$^{2-}$ and $1/4$ are vacant sites. For $A'_2$NiO$_2B_2$
\green{in the G and H$_2$ families}, 
a half of $A'$ sites are occupied by a cation with a valence of $2+$ and the other half are occupied by a $3+$ cation (e.g. Ba$_{0.5}$La$_{0.5}$), or $5/6$ are occupied by a $3+$ cation and $1/6$ are vacant sites.
}
\label{crystal_structures}
\end{center}
\end{figure*}

In this section, we discuss how we generate new layered nickelate structures that can potentially realize an ideal $d^9$ configuration where only the Ni 3\xx band forms the Fermi surface. 
To achieve the structure generation in a systematic and controlled way, we refer to the work of Tokura and Arima for high $T_c$ cuprates~\cite{Tokura_1990}, where they introduced the concept of ``Block Layer'' (BL) for classifying and designing crystal structures of cuprates. According to their classification, there are 14 types of BLs.
Among them, 10 types of BLs contain apical oxygen.
However, the valence of Ni cannot be monovalent when it is surrounded by six oxygen atoms.
Therefore, we hereafter consider the other 4 BLs that are free from apical oxygen and use their appropriate variants as a building block of layered nickelates. The considered BLs are the V-, G-, H$_1$- and H$_2$-type BLs in Ref.~\onlinecite{Tokura_1990}, whose crystal structures are shown in Fig.~\ref{block_layer}. Here, the gray and black spheres in the rectangular box represent cations and anions, respectively 
(see Table~\ref{table:list_of_nickelate} for the choice of cation and anion elements). 
The top and bottom planes correspond to the NiO$_2$ sheet. While the position of Ni is the same for the top and the bottom for the V and H$_1$ blocks (Figs.~\ref{block_layer}(a) and (b), see also Figs.~\ref{crystal_structures}(a),(b) and (c)),
there is a ``phase shift'' for the G and H$_2$ blocks (Figs.~\ref{block_layer}(c) and (d), see also Figs.~\ref{crystal_structures}\red{(d),(e) and (f)}).

To realize an ideal $d^9$ configuration of Ni, we need to prevent the BL band from 
making the Fermi surface.
Since the energy level of the Ni 3\xx orbital is close to the conduction band, 
the cation elements in the BL should be chosen carefully so that the energy level of the BL band becomes sufficiently high.
Therefore, 
for the cations in BLs, we use elements 
in the 1--3 groups (such as Sr and La), which strongly favors closed shell. 
This is in stark contrast with the case of the high-$T_c$ cuprates, where elements in the 11--13 groups (such as Hg and Tl) can be used as cations in the BLs thanks to the larger work function of Cu$^{2+}$.

\section{Crystal structure of candidate materials}
\label{sec_structure}

\begin{table*}[htbp]
{\renewcommand{\arraystretch}{1.5}
\caption{List of layered nickelates explored in this study.}
\label{table:list_of_nickelate}
\begin{ruledtabular}
\begin{tabular}{ccccc}
Family & Block layer & Space group & Composition & Structure \\ \hline
112 & V         & $P4/mmm$ & $A$NiO$_{2}$ ($A=$La, Nd) & Fig.~\ref{crystal_structures}(a) \\
1214 & H$_{1}$    & $P4/mmm$ & $AB_{2}$NiO$_{4}$ ($A=$ Li, Na, K, Rb, Cs; $B=$ Y, Sc, La, Nd)      & Fig.~\ref{crystal_structures}(b) \\
1213 & H$_{1}$    & $P4/mmm$ & $AB_{2}$NiO$_{3}$ ($A=$ Li, Na, K, Rb, Cs; $B=$ Mg, Ca, Sr, Ba, Yb) & Fig.~\ref{crystal_structures}(c) \\
 \multirow{2}{*}{G} & \multirow{2}{*}{G} & \multirow{2}{*}{$I4/mmm$} & $A_{2}$NiO$_{2}B^{\prime}_{2}$ ($A=$ Y, La, Nd, Lu; $B^{\prime}=$ O$_{0.5}$F$_{0.5}$) & Fig.~\ref{crystal_structures}(d) \\
  & & & $A'_{2}$NiO$_{2}B_{2}$ ($A'=$Ba$_{0.5}$La$_{0.5}$; $B=$ F, Cl, Br, I)  &  Fig.~\ref{crystal_structures}(e)\\
 H$_{2}$& H$_{2}$ & $I4/mmm$ & $A'_{2}$NiO$_{2}B_{2}$ ($A'=$Ba$_{0.5}$La$_{0.5}$; $B=$ F, Cl, Br, I) & Fig.~\ref{crystal_structures}(f) \\
\end{tabular}
\end{ruledtabular}
}
\end{table*}

Let us now assemble the NiO$_2$ layer and the BLs considered in the previous section. As is listed in Table~\ref{table:list_of_nickelate}, there are five possible families. For the V-type BL in high-$T_c$ cuprates, divalent cations such as Ca$^{2+}$ have been considered. In the case of nickelates, the cation should be trivalent to make the valence of the NiO$_2$ layer be $3-$. Indeed, the 112 family, which has already been synthesized experimentally~\cite{Hayward_1999,Hayward_2003,Crespin_2005,Kawai_2009}, contains a V-type BL consisting of La$^{3+}$ or Nd$^{3+}$. Since the work function of these elements is not sufficiently small,  
the NiO$_2$ layer in the \red{112} family is self-doped and electrons in the BL form Fermi pockets around \red{the $\Gamma$ and A points}.

Regarding the H$_1$-type BL in high-$T_c$ cuprates, a halogen layer sandwiched by two layers of divalent cations such as \red{Sr$^{2+}$ and Ba$^{2+}$} has been considered. However, the formation of a stable square CuO$_{2}$ plane has not been reported for the H$_{1}$-type BLs, with which lower-symmetry structures are favored~\cite{Boehlke_19990}.
Here, to improve the stability of layered nickelates with the H$_{1}$-type BLs, we consider \green{two variants} of the original H$_{1}$ structure.
In $AB_2$NiO$_4$ (\red{1214 family}, Fig.~\ref{crystal_structures}(b)), we replace the divalent cation with a trivalent cation $B$, and the halogen layer is replaced with an ($A$O$_2$)$^{3-}$ layer with $A$ being a monovalent cation. In $AB_2$NiO$_3$ (1213 family, Fig.~\ref{crystal_structures}(c)), we consider divalent cation $B$, and the halogen layer is replaced with \red{an ($A$O)$^{-}$} layer.

The G-type block can be seen in the mother compound of the $n$-type cuprate superconductor with the T'-structure~\cite{Tokura_1989}. The multilayer nickeates Nd$_4$Ni$_3$O$_8$~\cite{Lacorre_1992,Retoux_1998,Poltavets_2007} and La$_3$Ni$_2$O$_6$~\cite{Poltavets_2006} can be regarded as a stacked heterostructure of 
\green{the V-type and G-type blocks}.
While the valence of the G-type block ($Ln_2$O$_2$) is $2+$, we need a BL with a valence of $3+$. Thus we consider the following two variants of the G-type block. One is $A_2B^\prime_2$, where $A$ is a trivalent cation, such as Y, La, Nd, Lu, and $B^\prime$ is an anion with a valence of $1.5-$ (see Fig.~\ref{crystal_structures}(d)). This situation can be realized when $B^\prime$ is O$_{0.5}$F$_{0.5}$ or O$_{0.5}$H$_{0.5}$. Another possibility is $B^\prime=$O$_{0.75}\square_{0.25}$, i.e., the situation where 25\% of the $B^\prime$-sites are vacant. The other variant is $A'_2 B_2$, where $A'$ is a cation with a valence of 2.5+ and $B$ is a halogen such as F, Cl, Br and I (see Fig.~\ref{crystal_structures}(e)). 
\blue{
The valence of 2.5+ can be realized by mixing the divalent and trivalent cations with the ratio 1:1. 
Another possibility is that $5/6$ of the $A'$ sites are occupied by a $3+$ cation and $1/6$ are vacant.
In the present study, }
for $A'$, we consider the case of Ba$_{0.5}$La$_{0.5}$.

Finally, let us consider the possibility of the H$_2$-type family.
Again, while the H$_2$-type BL in high-$T_c$ cuprate is divalent, we need a trivalent BL for nickelates.
Figure~\ref{crystal_structures}(f) shows the crystal structure of $A_2^\prime$NiO$_2B_2$, 
where $A'$ is a cation with a valence of 2.5+. 
As in the case of \green{the G-family}, we consider  Ba$_{0.5}$La$_{0.5}$ as $A'$.

\section{Dynamical stability}
\label{sec_phonon}

\begin{figure*}[ptb]
\centering
\includegraphics[width=0.99\textwidth]{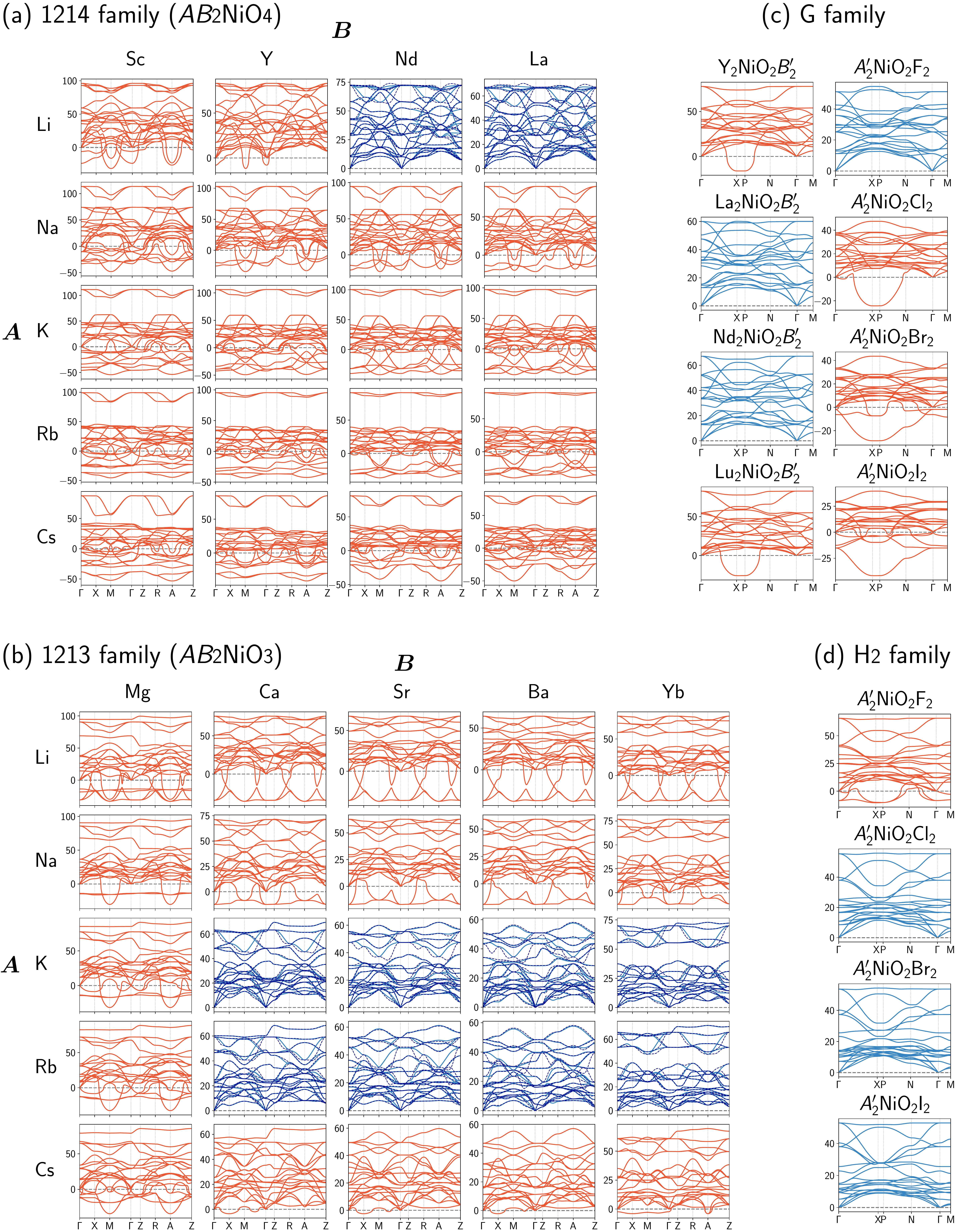}
\caption{Phonon dispersion curves of the (a) 1214-family, (b) 1213-family, (c) G-family, and (d) H$_{2}$-family nickelates calculated with the PBEsol functional. 
We take $A'$ and $B'$ atoms to be 
Ba$_{0.5}$La$_{0.5}$ and O$_{0.5}$F$_{0.5}$, respectively.
The unit of the vertical axes is meV. For the 1214 and 1213 families, the results based on the 2$\times$2$\times$2 and 4$\times$4$\times$2 supercells are shown with solid and dashed lines, respectively. Different colors are used for dynamically stable (blue) and unstable (red) compositions.}
\label{fig:phonons_nickelate}
\end{figure*}

In the previous section, we have systematically generated 57 new layered nickelates following our design principles.
To assess the dynamical stability of these 57 layered nickelates of Table \ref{table:list_of_nickelate}, here we perform \textit{ab initio} phonon calculations of these structures. The calculated phonon dispersion curves of 57 nickelates are shown in Fig.~\ref{fig:phonons_nickelate}, where the curves of 16 dynamically stable materials are shown with blue lines.
    
Among the explored 1214-family nickelates, only LiNd$_{2}$NiO$_{4}$ and LiLa$_{2}$NiO$_{4}$ are predicted to be dynamically stable, as shown in Fig.~\ref{fig:phonons_nickelate}(a). 
For the $A$ site cation, Li$^{1+}$ is preferable because its ionic radius is the closest to that of Ni$^{1+}$, resulting in a good matching of the in-plane lattice constant between the NiO$_{2}$ and LiO$_{2}$ layers (see Fig.~\ref{crystal_structures}(b)). With increasing the $A$ site ionic radius, the instability of phonon modes enhances due to the increase in the lattice mismatch. When $A$ = Li, the structure becomes dynamically stable if the $B$ site cation is either Nd or La. This result is reasonable because the ionic radii of Nd$^{3+}$ and La$^{3+}$ are similar.
Since the ionic radius of Y$^{3+}$ is also similar to that of Nd$^{3+}$ but somewhat smaller,
LiY$_{2}$NiO$_{4}$ has unstable phonon modes around the $\Gamma$ and M points. These soft modes respectively correspond to the distortion of the oxygen atoms in the in-plane and out-of-plane directions. Since the instability of these soft modes is rather weak, applying tensile or compressive strain to LiY$_{2}$NiO$_{4}$, for example, by growing it on a substrate, may stabilize the $P4/mmm$ structure.

Figure \ref{fig:phonons_nickelate}(b) shows the calculated phonon dispersion curves of the 1213-family nickelates. The dynamical stability is realized when $A$ = K or Rb and $B$ = Ca, Sr, Ba, or Yb. In contrast with the 1214 family, a relatively large cation is preferred for the $A$ site in the 1213 family because the number of atoms in the $A$O layer is less than that of the NiO$_{2}$ layer. The structures with Cs become unstable probably because the ionic radius of \red{Cs$^{+}$} is too large for the $A$ site.
When $A$ = K or Rb, the structure is stable if $B$ site is either Ca, Sr, Ba, or Yb. 
The ionic radii of Ca$^{2+}$ and Sr$^{2+}$ are close to those of Nd$^{3+}$ and La$^{3+}$ which are used as the V block in the stable 112 family. When $B$ = Mg, the structure becomes unstable irrespective of types of the $A$ site cation, which can be attributed to the small ionic radius of Mg$^{2+}$. Also, the large electronegativity of Mg, which is larger than that of the other alkaline earth elements, tends to hamper the formation of a stable Mg$^{2+}$ cation.

Among the studied four G-family $A_{2}$NiO$_{2}B'_{2}$ nickelates, La$_{2}$NiO$_{2}B'_{2}$ and Nd$_{2}$NiO$_{2}B'_{2}$ are predicted to be dynamically stable, and the other two compositions are unstable. This result is again reasonable considering that LaNiO$_{2}$ and NdNiO$_{2}$ have been synthesized, whereas the other two have not been reported so far. Since the ionic radii of Y$^{3+}$ and Lu$^{3+}$ are somewhat smaller than that of Nd$^{3+}$, the phonon mode around the X and P points, which involves in-plane distortion of the oxygen and \red{$B^{\prime}$ atoms}, become unstable and the $I4/mmm$ structure changes into a lower-symmetry structure.

Next, we discuss the dynamical stability of the G-family (T'-type structure) and H$_{2}$-family (T-type structure) $A'_{2}$NiO$_{2}B_{2}$ nickelates. As can be seen in the right panels of Fig.~\ref{fig:phonons_nickelate}(c) and the panels of Fig.~\ref{fig:phonons_nickelate}(d), the dynamically stable structure changes clearly depending on the halogen atoms; the G-type structure is stable only when $B$ = F, while the H$_{2}$-type structure is stable when $B$ = Cl, Br, or I. 
Interestingly, our computational prediction is consistent with the experimental findings on the related compounds. For example, Sr$_{2}$CuO$_{2}$F$_{2}$ and Ba$_{2}$PdO$_{2}$F$_{2}$ crystallize in the G-type structure~\cite{Kissick_1997,Baikie_2003}, whereas Sr$_{2}$CuO$_{2}$Cl$_{2}$ and Ba$_{2}$PdO$_{2}$Cl$_{2}$ display the H$_{2}$-type structure~\cite{Miller_1990,Hiroi_1994,Tsujimoto_2014}. We can infer from these consistent results that the H$_{2}$-type structure tends to become more stable than the G-type structure with increasing the halogen ion radius.

Our systematic \textit{ab initio} phonon calculations predict that 16 compositions out of 57 are dynamically stable. The dynamically stable compositions are LiNd$_{2}$NiO$_{4}$, LiLa$_{2}$NiO$_{4}$, $AB_{2}$NiO$_{3}$ with $A$ = (K, Rb) and $B$ = (Ca, Sr, Ba, Yb), La$_{2}$NiO$_{2}B'_{2}$ and Nd$_{2}$NiO$_{2}B'_{2}$ where $B'$ = O$_{0.5}$F$_{0.5}$, G-family $A'_{2}$NiO$_{2}$F$_{2}$, and H$_{2}$-family $A'_{2}$NiO$_{2}B_{2}$ ($B$ = Cl, Br, I) with $A'$=Ba$_{0.5}$La$_{0.5}$. Although the dynamical stability does not guarantee the synthesizability of materials, dynamically stable materials are, at least, metastable. Indeed, metastable phases appear ubiquitously~\cite{Sun_2016}. Therefore, we believe that the predicted metastable nickelates can be synthesized by using experimental techniques, including the high-pressure synthesis and the pulsed laser deposition
\footnote{
\rr{It is still a formidable task to perform a convex hull calculation for quaternary compounds simply because the search space is too vast to explore. Thus, we leave it as an interesting and challenging future problem}
}
.

\section{Electronic structure}
\label{sec_band}

\begin{figure*}[htbp]
\vspace{0cm}
\begin{center}
\includegraphics[width=0.99\textwidth]{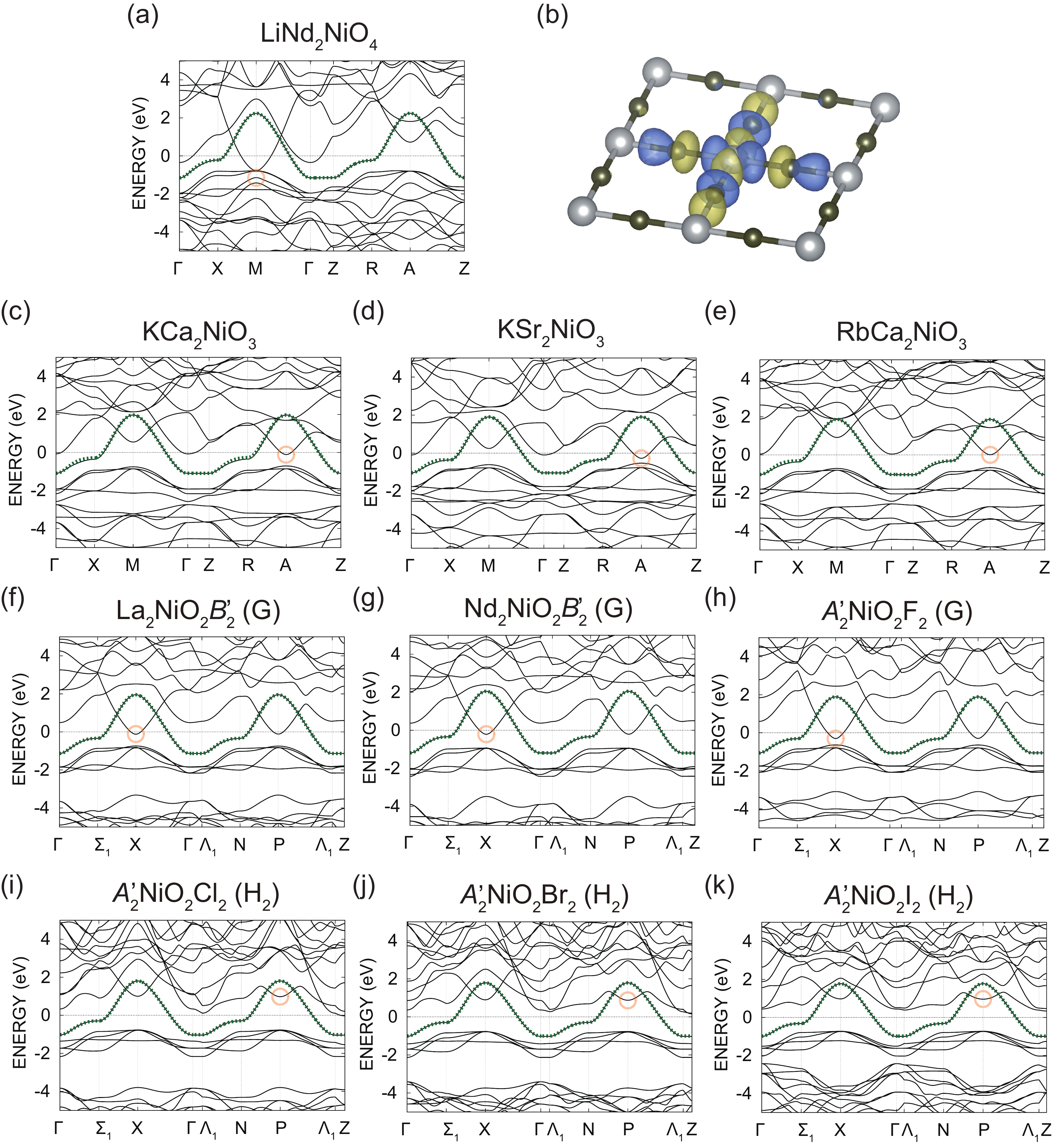}
\caption{
(a) Electronic band structure of LiNd$_2$NiO$_4$.
(b) Wannier function of the Ni $d_{x^2-y^2}$ orbital in LiNd$_2$NiO$_4$.
(c)--(k) Electronic band structures of KCa$_2$NiO$_3$, KSr$_2$NiO$_3$, RbCa$_2$NiO$_3$, La$_2$NiO$_2B'_2$, Nd$_2$NiO$_2B'_2$, 
$A'_2$NiO$_2$F$_2$, $A'_2$NiO$_2$Cl$_2$, $A'_2$NiO$_2$Br$_2$, and $A'_2$NiO$_2$I$_2$, respectively.
The energy is measured from the Fermi level.
\rr{
To compare the band structures among the different
families with different space group, we employ the consistent $\bm{k}$ path: The $\Gamma$, $\Sigma_1$, X, $\Lambda_1$, N, and P points in the panels (f-k) corresponds to $\Gamma$, X, M, Z, R and A points in the panels (a,c-e), respectively.
More specifically,
}
\red{$\Sigma_1$ and $\Lambda_1$ in the $I4/mmm$ structures are (0.5,0,0) and (0,0,0.5) in the conventional basis of the conventional unit cell shown in Fig.~\ref{crystal_structures}, respectively.}
\rr{
Since $k_z=0$ and $k_z=\pi$ are on the same plane in the reciprocal space in $I4/mmm$, we show a band at $k_z=\pi/2$ instead of $k_z=\pi$.
}
The open circles indicate the band minimum of the bonding band formed by the interstitial $s$ state and the \xy state of the neighboring cation. The green dotted curves are the Wannier-interpolated band of the effective single-band model.
}
\label{Ni_band_all}
\end{center}
\end{figure*}

\begin{figure*}[htbp]
\centering
\includegraphics[width=0.65\textwidth]{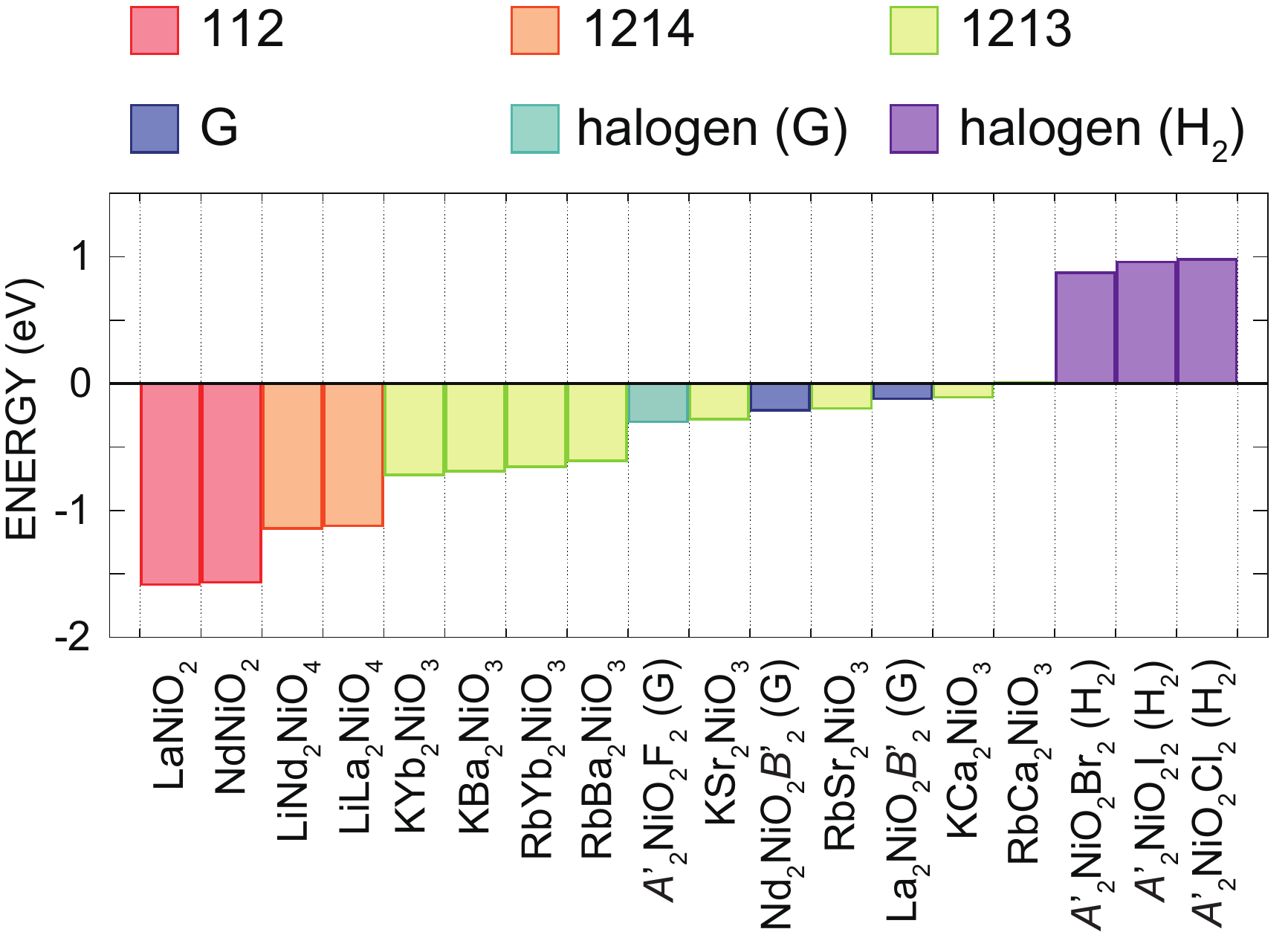}
\caption{
Lowest energy of the bonding band made from the interstitial $s$ and the cation $d_{xy}$ orbitals with respect to the Fermi level.
}
\label{sd_energy}
\end{figure*}

Let us now move on to the electronic structure of the compounds that are shown to be dynamically stable in the previous section. 
Here, we discuss how the position of the bonding band 
between the interstitial $s$ and the cation  $d_{xy}$ orbitals
that makes a large extra Fermi pocket  in the 112 family 
changes when adopting the different types of BLs. 
To this end, we show the electronic band structures calculated for 10 representative dynamically-stable systems in Fig.~\ref{Ni_band_all}. 
For each material, we indicate the position of the band minimum of the bonding band,
which occurs at $(k_x,k_y)=(\pi,\pi)$,
by open circles.
We see in the figure that the energy level of the bonding band is lifted from that of the 112 family for the all new nickelates. 

To see this point more quantitatively, we compare the energy levels of the bonding band in Fig.~\ref{sd_energy}. Intriguingly, the nickelates belonging to the H$_{2}$ family are the most promising in that the energy level of the bonding band is significantly higher than the Fermi level, 
and only the Ni 3\xx band makes the Fermi surface. 
See Table~\ref{Ene_sd} in Appendix~\ref{ene_table} for the detail of the onsite level of the interstitial $s$ and cation \xy orbitals, and the bandwidth of bonding and antibonding bands.
In what follows, we look into the detail of the electronic structure of the four families.

Figure \ref{Ni_band_all}(a) shows the electronic band structure of LiNd$_2$NiO$_4$. We also show the Wannier function originating from \red{the Ni 3\xx orbital} in Fig.~\ref{Ni_band_all}(b), which is the basis of the effective single-orbital model discussed in Sec.~\ref{sec_effective}. The bottom of the bonding band made from the interstitial $s$ and \red{the Nd $5d_{xy}$ orbitals} (indicated by the orange circle) is higher than that in the 112 family by 450 meV (see also Fig.~\ref{sd_energy}). However, this bonding band still makes rather large Fermi pockets.

In the 1213 family, the trivalent ions (i.e., La$^{3+}$ and Nd$^{3+}$) used in the 1214 family are replaced with the divalent ions (i.e., Ca$^{2+}$ and Sr$^{2+}$). 
Then, the energy level of the cation \xy orbital becomes higher, which makes the energy of the bonding state higher.
Also, because the 3$d_{xy}$ orbital in Ca$^{2+}$ and 4$d_{xy}$ orbital in Sr$^{2+}$ are spatially more localized than 5$d_{xy}$ orbital in La$^{3+}$ and Nd$^{3+}$,
the bandwidth of the bonding and antibonding bands becomes small, resulting in a smaller Fermi pocket. 
Indeed, in the 1213 family, the BL makes almost no extra Fermi pocket (see Figs.~\ref{Ni_band_all}(c)--(e) and \ref{sd_energy}). 
In particular, in RbCa$_2$NiO$_3$, the extra Fermi pocket does not exist, so that the Fermi surface originates only from \red{the Ni 3\xx orbital}.

We show the electronic band structures of the G-family \red{La$_2$NiO$_2B'_2$, Nd$_2$NiO$_2B'_2$, and $A'_2$NiO$_2$F$_2$} in Figs.~\ref{Ni_band_all}(f), (g), and (h), respectively. 
\blue{ 
The size of the Fermi pocket formed by the bonding band is again smaller compared to the 112 family. 
In the case of the 112 family, 
the interstitial $s$ orbital is surrounded by cations. 
On the other hand, in the G family,
\green{the negatively charged halogen site}
is located near the interstitial $s$ orbital, 
which makes the energy of 
the interstitial $s$ higher.} 
\blue{In La$_2$NiO$_2B'_2$ and Nd$_2$NiO$_2B'_2$, this leads to large energy difference between the interstitial $s$ and $d_{xy}$ orbitals, 
and the hybridization between the two orbitals becomes smaller. 
Therefore, the bonding-antibonding splitting is much smaller than that of \NNO, and consequently, the Fermi pocket formed by the bonding band becomes small. 
In \green{the} case of $A'_2$NiO$_2$F$_2$ with $A'=$ Ba$_{0.5}$La$_{0.5}$, 
the energy level of the $5d_{xy}$ orbital also rises because of the change of the valence of the cation from 3+ to 2.5+. 
This also leads to a smaller Fermi pocket even though the bonding-antibonding splitting is comparable to that of the 112 family.
}

We also show the electronic band structures of \red{$A'_2$NiO$_2$Cl$_2$, $A'_2$NiO$_2$Br$_2$, and $A'_2$NiO$_2$I$_2$} belonging to the H$_2$ family (Figs.~\ref{Ni_band_all}(i), (j) and (k)).
These materials have no Fermi pocket originating from the bonding band. 
The energy level of the interstitial $s$ orbital in the H$_2$ family is very high. 
This is because the negatively-charged halogen site
is located just on top of the interstitial site; thus the increase in the onsite level of the interstitial $s$ orbital is even larger than that of the G family.

The energy minimum of the bonding state formed by the interstitial $s$ and $d_{xy}$ orbitals is summarized in Fig.~\ref{sd_energy}.
When the energy minimum is lower, the size of the Fermi pocket made by the bonding band is larger. We see that the energy minimum is highest for the H$_2$ family and lowest for LaNiO$_2$. Our results indicate that the H$_2$ family would provide a firm ground to study the possibility of high-$T_c$ superconductivity in the Mott-Hubbard regime of the Zaanen-Sawatzky-Allen classification~\cite{Zaanen_1985}.

\section{Effective Single-Orbital Model}
\label{sec_effective}

\begin{table*}[bt] 
\caption{
Hopping and interaction parameters in the single-orbital model. 
$t$, $t'$, $t''$ are the nearest, next-nearest, and third-nearest hopping integrals, respectively. 
$U$ is the onsite Hubbard interaction. 
The energy unit is eV.
The data for NdNiO$_2$ are taken from Ref.~\onlinecite{Nomura_arXiv}.
} 
\vspace{0.2cm}
\begin{tabular}{@{\ \ \ \ }c@{\ \ \ \ \  }c@{\ \ \ \ \ \ \ }c@{\ \ \ \ \ }c@{\ \ \ \ \ }c@{\ \ \ \ \ }c@{\ \ \ \ \ }c@{\ \ \ \ \ }c@{\ \ \ \ }}
\hline \hline \\ [-8pt]   
 Composition & Family & $t$  & $t'$ & $t''$ & $U$ & $|U/t|$ & $t'/t$  \\ [+1pt]
\hline \\ [-8pt] 
NdNiO$_2$       & 112  & $-0.370$ & 0.092 & $-0.045$ & 2.608  & 7.052 & $-0.250$ \\ [+1pt]
LiNd$_2$NiO$_4$ & 1214 & $-0.404$ & 0.099 & $-0.052$ & 2.865 & 7.086 & $-0.245$ \\ [+1pt]
KCa$_2$NiO$_3$  & 1213 & $-0.370$ & 0.104 & $-0.049$ & 3.222 & 8.701 & $-0.280$  \\ [+1pt]
KSr$_2$NiO$_3$  & 1213 & $-0.357$ & 0.102 & $-0.048$ & 2.985 & 8.366 & $-0.285$  \\ [+1pt]
RbCa$_2$NiO$_3$ & 1213 & $-0.352$ & 0.100 & $-0.046$ & 3.347 & 9.522 & $-0.284$ \\ [+1pt]
Nd$_2$NiO$_2B'_2$     & G & $-0.389$ & 0.098 & $-0.049$ & 3.241 & 8.335 & $-0.253$ \\ [+1pt]
$A'_2$NiO$_2$F$_2$ & G & $-0.355$ & 0.095 & $-0.045$ & 3.399 & 9.587 & $-0.269$  \\ [+1pt]
$A'_2$NiO$_2$Br$_2$& H$_2$ & $-0.337$  &  0.089 &  $-0.039$ & 3.586 & 10.637 & $-0.263$  \\ [+1pt]
\hline \\ [-8pt] 
RbSr$_2$PdO$_3$  & 1213 & $-0.493$  & 0.116  & $-0.076$   & 2.362 & 4.795  &  $-0.236$ \\ [+1pt]
$A'_2$PdO$_2$Cl$_2$  & G & $-0.443$  & 0.097  & $-0.063$  & 2.699 & 6.085 &  $-0.218$ \\ [+1pt]
\hline \hline 
\end{tabular}
\label{model} 
\end{table*} 

As we have discussed in the previous sections, our proposed materials are closer to an ideal $d^9$ system compared to \NNO. 
Then, the electronic structure around the Fermi level is well described by the single-orbital model in the Mott-Hubbard regime. 
Here, we construct effective single orbital models for \green{the} $d^9$ candidate nickelates
by using a combination of the maximally localized Wannier functions~\cite{Marzari_1997,Souza_2001} and 
the constrained random phase approximation (cRPA)~\cite{Aryasetiawan_2004}.
The detail of the calculations can be found in Appendix~\ref{Appendix_cRPA}.

Table~\ref{model} shows the hopping and interaction parameters of the single-orbital model. 
The band dispersion curves obtained by the constructed tight-binding Hamiltonians well reproduce those of the
\rr{DFT}
calculations (see Fig.~\ref{Ni_band_all}).
The property of the single-orbital model is summarized as follows:
\begin{itemize}
    \item All the materials are in the strongly correlated regime: $|U/t|$ is around 7--11.  
    Correlation strength is comparable to that of the cuprates~\cite{Tadano_2019,Hirayama_2018,Hirayama_2019,Werner_2015,Jang_2016,Nilsson_2019,Nomura_arXiv}.  
    \rr{Therefore, in the \rrr{ideal $d^9$ configuration}, the ground state will be an antiferromagnetic Mott insulator similarly to the cuprates, in contrast with the existing NdNiO$_2$, where the Mott insulating behavior is masked by the self-doping.}
    \item The values of frustration parameter $t'/t$ lie between $-0.29$ and $-0.24$, which are also comparable to those of the  cuprates~\cite{Botana_arXiv,Nomura_arXiv,Pavarini_2001}. 
    \item Materials with large in-plane lattice constant tends to have small hopping integral $t$.  
    \item When the cation-layer bonding orbital forms a Fermi pocket, it contributes to the screening of the Coulomb interaction, although the effect is not drastic to make the system weakly correlated~\cite{Nomura_arXiv}.  
    Therefore, when such a Fermi pocket is small or eliminated, the Hubbard $U$ tends to become large. 
    In particular, the H$_2$-family nickelates have the most isolated Ni 3\xx band; therefore, they have the strongest electron correlations. 
  \end{itemize}


\section{Materials design of palladate analog of high $T_c$ cuprates}
\label{sec:discussion}

\rr{Let us finally extend the above discussion to a systematic materials design of layered palladates.
In fact, the electronic structure of the layered palladium oxides La$_2$PdO$_4$, LaPdO$_2$, and La$_4$Pd$_3$O$_8$ have been studied via first-principles calculation~\cite{Botana18}.
While the filling of the Pd $4d$ orbitals deviates from $d^9$ in these palladates, here we seek the possibility of $d^9$ palladates. 
}
The correlation strength would be weaker than that of nickelates; hence, by combining nickelates and palladates, we might be able to study a wide range of the $|U/t|$ parameter region.

To seek dynamically stable $d^9$ layered palladates, we first optimized the structure parameters of the 57 palladates and then performed systematic phonon calculations. 
See Appendix \ref{sec:structure_and_phonon_Pd} for the details of the methods.
We see in Fig.~\ref{fig:Palladate_phonon} that only 6 compositions out of 57 are predicted to be dynamically stable. The stable compositions are $A$Sr$_{2}$PdO$_{3}$ ($A$ = K, Rb, Cs), CsBa$_{2}$PdO$_{3}$, and G-type $A'_{2}$PdO$_{2}B_{2}$ ($B$ = F, Cl). 
Because Pd has a larger atomic radius than Ni, 
the structure tends to be stable when the elements in the BLs have a larger atomic radius compared to those in nickelates.

\begin{figure*}[htbp]
\vspace{0cm}
\begin{center}
\includegraphics[width=0.99\textwidth]{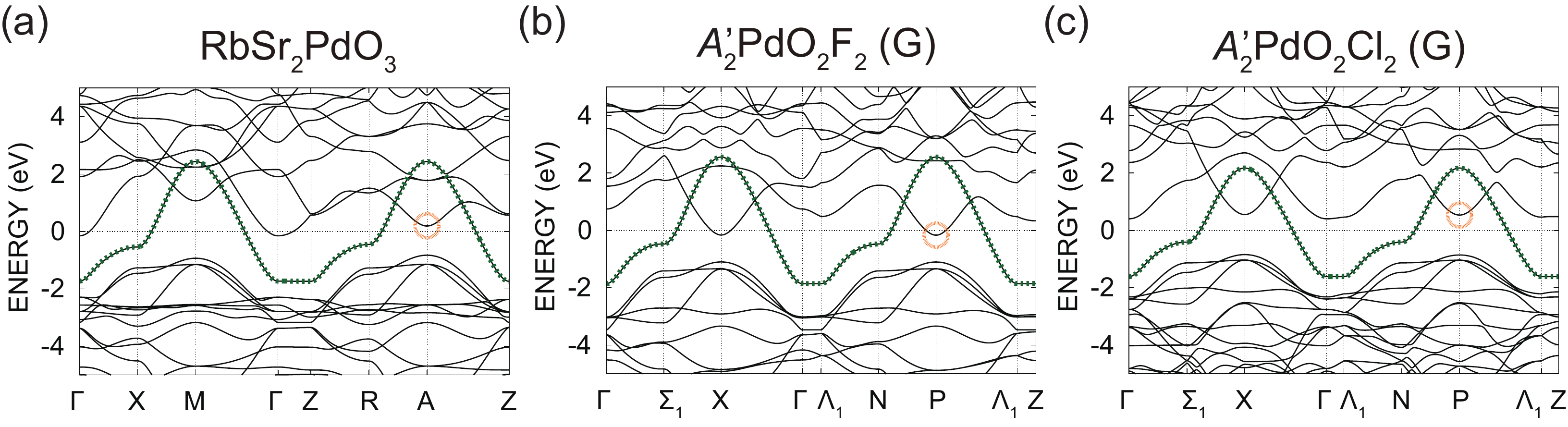}
\caption{
(a)--(c) Electronic band structures of RbSr$_2$PdO$_3$, $A'_2$PdO$_2$F$_2$, and $A'_2$PdO$_2$Cl$_2$, respectively.
The energy is measured from the Fermi level.
The open circles indicate the band minimum of the bonding band formed by the interstitial $s$ state and the \xy state of the neighboring cation. The green dotted curves are the Wannier-interpolated band of the effective single-band model.
}
\label{Pd_band}
\end{center}
\end{figure*}

The electronic band structures of RbSr$_{2}$PdO$_3$ in the 1213 family,  $A'_2$PdO$_2$F$_2$ and $A'_2$PdO$_2$Cl$_2$ in the G family are shown in Figs.~\ref{Pd_band}(a), (b), and (c), respectively.
The band structure of RbSr$_{2}$PdO$_3$ and $A'_2$PdO$_2$Cl$_2$ have no extra Fermi surface originating from the interstitial $s$ and $d_{xy}$ orbitals.
The bandwidth of the Pd compounds is larger than that of the Ni compounds as expected.

As in the nickelates, we construct effective single-orbital models for the palladates.
The Wannier interpolated bands can be found in Fig.~\ref{Pd_band}. 
The hopping and interaction parameters are shown in Table~\ref{model}.
We see that: 
\begin{itemize}
    \item  The kinetic energy scale is larger than that of nickelates. 
    \item  $|U/t|$ in the palladates is smaller compared to that of nickelates because of the extended $4d$ orbitals making $U$ smaller and $t$ larger. Hence, the correlation strength is weaker. $|U/t|$ is around 5--6.
    \item Material dependence of the correlation strength is the same as that of nickelates; when the Fermi pocket formed by the cation-layer bonding state is small or eliminated, the Hubbard $U$ tends to become large.  
    In particular, the Hubbard $U$ in the G-type $A'_{2}$PdO$_2$Cl$_2$ is comparable to that of NdNiO$_2$. 
  \end{itemize}

\section{Summary}
\label{sec_conclusion}
We performed a systematic computational search for layered nickelates that are dynamically stable and whose electronic structure better mimics that of high-$T_c$ cuprates than Sr-doped \NNO.
While there is 
\blue{
\green{more than 10\%}
self-doping of the Ni 3\xx orbital }
in \NNO, by referring to the work by Tokura and Arima~\cite{Tokura_1990} for high-$T_c$ cuprates, and by performing comprehensive phonon calculations, we found dynamically-stable layered nickelates where the self-doping is almost or completely suppressed. 
We derived effective single-band models for the promising materials and found that they are in the strongly correlated regime: $|U/t|$ is around 7--11. By replacing Ni with Pd, we also designed promising palladate analogs of high-$T_c$ cuprates. Once synthesized, these materials will provide a firm ground for studying the possibility of high-$T_c$ superconductivity in the Mott-Hubbard regime of the Zaanen-Sawatzky-Allen classification~\cite{Zaanen_1985}.

\begin{acknowledgments}
We are grateful to Y. Tokura, T. Katase, M. Uchida, Y. Taguchi, S. Ishiwata, and S. Seki for their helpful comments and to K. Nakamura and Y. Yoshimoto for fruitful discussion. We acknowledge the financial support of JSPS Kakenhi Grant No. 16H06345 (MH, TT, YN, and RA), No. 17K14336 (YN), No. 18H01158 (YN), and No. 19H05825 (RA). The figures of the crystal structures are created by the VESTA software~\cite{Momma_2011}.
\end{acknowledgments}
MH and TT contributed equally to this work.

\appendix
\section{Details of computational methods}
\label{sec_methods}
\rr{
  For efficiently performing the different types of calculations, we employed several different DFT and GW codes, including the Vienna Ab initio Simulation Package (\textsc{vasp})~\cite{kresse_1996}, \QE (QE)~\cite{QE-2017}, and OpenMX~\cite{Ozaki_2003}. The details of each calculation are described below. 
}

\subsection{Structural optimization}
\label{subsec_structure_opt}
The structure optimization of the 1214- and 1213-family nickelates was performed with \textsc{vasp}~\cite{kresse_1996}. We used the projector augmented wave (PAW)~\cite{Blochl_1994,Kresse_1999} potential sets recommended by \textsc{vasp} and set the kinetic energy cutoff to 600 eV for systems containing Li and 500 eV for the others. 
For the G- and H$_{2}$-family nickelates, the structure optimization was conducted by using QE~\cite{QE-2017} with the optimized norm-conserving Vanderbilt (ONCV) pseudopotentials~\cite{Hamann_2013} adopted from the PseudoDojo~\cite{Setten_2018}. Mixed pseudopotentials of Ba$_{0.5}$La$_{0.5}$ and O$_{0.5}$F$_{0.5}$, generated based on the virtual crystal approximation (VCA), were used for representing the cation $A'$ and the anion $B'$, respectively. The kinetic cutoff energy of 100 Ry was employed in the QE calculations.
For both \textsc{vasp} and QE calculations, the $\bm{k}$ point mesh was generated automatically in such a way that the mesh density in the reciprocal space becomes larger than 450 \AA$^{-3}$. 
Also, $4f$ electrons of Nd and Lu were frozen in the core.
Since phonon frequencies are rather sensitive to the lattice parameters, we employed the generalized-gradient approximation (GGA) with the Perdew-Burke-Ernzerhof parameterization for solid (PBEsol)~\cite{Perdew_2008}, which is reported to work well for predicting lattice parameters of oxides and other solids~\cite{Wahl_2008,Csonka_2009}. The complete information of the optimized crystal structures is summarized in Appendix \ref{sec:structure_and_xray} and \ref{sec:structure_and_phonon_Pd}.

\subsection{Phonon calculation}
\label{subsec_phonon}
The phonon calculations of the 57 layered nickelates were performed by computing second-order interatomic force constants with sufficiently large supercells. For quickly identifying dynamically stable compositions of the 1214 and 1213 families, we started the phonon calculations with 2$\times$2$\times$2 supercells, which contain 64 (56) atoms for the 1214 (1213) family. For each nickelate, the dynamical stability is assessed from the presence of unstable phonon modes; if an unstable phonon mode ($\omega_{\bm{q}\nu}^{2}<0$) exists on the commensurate 2$\times$2$\times$2 $\bm{q}$ points, the composition is dynamically unstable and therefore excluded from the candidate list. After the first quick screening based on 2$\times$2$\times$2 supercells, we performed a second screening based on 4$\times$4$\times$2 supercells for confirming the dynamical stability of the candidate compositions.
\rr{
As can be seen in Fig.~\ref{fig:phonons_nickelate}, the supercell size dependence of the phonon dispersion curves is minor in the low-frequency region, indicating that the 4$\times$4$\times$2 supercell calculation is adequate for discussing dynamical stability.
}
Since the number of explored composition is relatively few for the G- and H$_{2}$-family, we only employed 3$\times$3$\times$1 conventional cells, which contain 126 atoms and therefore are sufficiently large, for judging the dynamical stability of the G- and H$_{2}$-type nickelates. 
All of the phonon calculations were conducted by using \textsc{alamode} software~\cite{Tadano_2014, Tadano_2018}.

\subsection{Band structure calculation}

The band structures and the maximally-localized Wannier functions (MLWFs) shown in Sec.~\ref{sec_band} are obtained by using the \textit{ab initio} code OpenMX~\cite{Ozaki_2003}.
We employed the valence orbital sets
Li8.0-$s3p3d2$, K10.0-$s4p3d2$, Rb11.0-$s3p3d2f2$, Cs12.0-$s3p2d2f2$,
Ca9.0-$s4p3d2$, Sr10.0-$s3p1d2f1$, Ba10.0-$s3p2d2f2$, Yb8.0-$s3p2d2f1$,
La8.0-$s3p3d3$, Nd8.0\_OC-$s2p2d2f1$,
Ni6.0H-$s4p3d2f1$, Pd7.0-$s2p2d2f1$,
O-5.0$s3p3d2$,
F6.0-$s2p2d1$, Cl7.0-$s2p2d1$, Br7.0-$s3p2d1$, I7.0-$s3p3d2f1$,
5.0-$s3p3d2$ for $N=8.5$ virtual atom, and 8.0-$s3p3d2$ for $N=56.5$ virtual atom.
The energy cutoff for the numerical integration was set to 150 Ry.
The Wannier functions of the interstitial $s$ and cation $d_{xy}$ orbitals were constructed from the outer energy window of [$-$2 eV: 12 eV] for $A$NiO$_2$ and $AB_2$NiO$_4$, [$-$1 eV: 12 eV] for $AB_2$NiO$_3$, $A_2$NiO$B'_2$, $AB_2$PdO$_3$, and $A'_2$PdO$_2$F$_2$, and [0 eV: 12 eV] for $A'_2$NiO$_2B_2$ and $A'_2$PdO$_2$Cl$_2$.
The Wannier functions of the $3d_{x^2-y^2}$ orbital for the Ni compounds were constructed from the outer energy window of [$-$1.5 eV: 2.5 eV], where we did not use the inner energy window for $A$NiO$_2$ and $AB_2$NiO$_4$ and used the inner energy window of [$-$0.4 eV: $-$0.3 eV] for KSr$_2$NiO$_3$, [$-$0.45 eV: $-$0.35 eV] for $A'_2$NiO$_2$F$_2$, and [$-$0.35 eV: $-$0.25 eV] for the other systems.
The Wannier functions of the $4d_{x^2-y^2}$ orbital for the Pd compounds were constructed from the outer energy window of [$-$2 eV: 2.5 eV] for KSr$_2$PdO$_3$ and $A'_2$PdO$_2$F$_2$, and [$-$2 eV: 3 eV] for $A'_2$PdO$_2$Cl$_2$, where we used the inner energy window of [$-$0.6 eV: $-$0.3 eV] for all the Pd-based compounds.
The lattice constants obtained by the \textsc{vasp} and QE calculations, shown in Tables \ref{table:structures_1214_PBEsol}--\ref{table:structures_H2type_Pd}, were used except for NdNiO$_{3}$ and LaNiO$_{3}$; for these 112-family nickelates, we used the experimental lattice constants~\cite{Hayward_2003,Crespin_2005}.
We employed the 12$\times$12$\times$12 $\bm{k}$ mesh for the 112 family, 12$\times$12$\times$8 $\bm{k}$ mesh for all other families.

\subsection{cRPA calculation}
\label{Appendix_cRPA}

The derivation of the single-band effective Hamiltonians shown in Secs.~\ref{sec_effective} 
and 
\ref{sec:discussion} was performed by using RESPACK~\cite{RESPACK_URL}\footnote{Algorithms and applications of RESPACK can be found in Refs.~\cite{Nakamura_2016,Nakamura_2009,Nakamura_2008,Nohara_2009,Fujiwara_2003}.}. 
First, we performed the DFT band structure calculations using QE~\cite{QE-2017}.    
We employed Perdew-Burke-Ernzerhof (PBE)~\cite{Perdew_1996} norm-conserving pseudopotentials generated by the code ONCVPSP (Optimized Norm-Conserving Vanderbilt PSeudopotential)~\cite{Hamann_2013}, which were downloaded from the PseudoDojo~\cite{Setten_2018}.
The kinetic energy cutoff was set to be 100 Ry for the wave functions, and 400 Ry for the charge density. 
We used $11\times 11 \times 7$ ${\bf k}$-mesh for 1213 and 1214 families and 
 $11\times 11 \times 11$ ${\bf k}$-mesh for G and $H_2$ families.

We constructed the maximally localized Wannier functions~\cite{Marzari_1997,Souza_2001} using RESPACK.
The outer energy window was set as follows: 
\begin{itemize}
    \item LiNd$_2$NiO$_4$ (1214):  [$-1.15$ eV : 2.25 eV]     
    \item KCa$_2$NiO$_3$ (1213):   [$-1.10$ eV : 2.00 eV]     
    \item KSr$_2$NiO$_3$ (1213):   [$-1.00$ eV : 1.95 eV]     
    \item RbCa$_2$NiO$_3$ (1213):  [$-1.05$ eV : 1.90 eV]      
    \item Nd$_2$NiO$_2B'_2$ (G):   [$-1.15$ eV : 2.10 eV]     
    \item $A'_2$NiO$_2$F$_2$ (G): [$-1.00$ eV : 1.95 eV]   
    \item $A'_2$NiO$_2$Br$_2$ (H$_2$): [$-1.05$ eV : 1.80 eV]   
    \item RbSr$_2$PdO$_3$ (1213):  [$-1.60$ eV : 2.55 eV]      
    \item $A'_2$PdO$_2$Cl$_2$ (G): [$-1.55$ eV : 2.20 eV]    
\end{itemize}
For the nickelate of the 1213, G, and H$_2$ families, we also used the inner window of [$-0.35$ eV : 0.25 eV]. 
For the palladates, the inner window was set to be [$-0.45$ eV : 0.35 eV].

The interaction parameters were calculated based on the cRPA method~\cite{Aryasetiawan_2004}, which is implemented in RESPACK. 
For the band disentanglement, we follow the scheme proposed in Ref.~\onlinecite{Sasioglu_2011}.
The energy cutoff for the dielectric function was set to be 20 Ry. 
The total number of bands used in the calculation of the polarization was 160.
The maximum energy of the unoccupied bands reaches at least 40 eV with respect to the Fermi level. 

\subsection{\rr{GWA calculation}}
\label{Appendix_GWA}
\rr{
The calculation of the electronic structure in the GWA is based on the full potential linearized muffin tin orbital implementation~\cite{Methfessel}.
The exchange correlation functional is obtained by the local density approximation (LDA) of the Ceperley-Alder type~\cite{Ceperley80}, and spin degree of freedom was neglected.
The LDA calculation was performed with the $12 \times 12 \times 12$ $\bm{k}$ mesh.
The adopted muffintin radii are as follows:
$R_{\text{La}}$= 2.90 bohr, 
$R_{\text{Ni}}$= 2.15 bohr, and $R_{\text{O}}$= 1.50 bohr.
The angular momentum of the atomic orbitals was taken into account up to $l=4$ for all the atoms.
}

\rr{
The GW calculations use a mixed basis
consisting of the products of two atomic orbitals and the interstitial plane wave~\cite{Schilfgaarde06}.
In the GW calculation, the $6 \times 6 \times 6$ {\bf k} mesh was employed.
By comparing the calculations with a smaller {\bf k} mesh, we confirmed that these conditions give well-converged results.
We included bands in [$-32.8$: $99.1$] eV
(84 bands) for the calculation of the screened interaction and the self-energy.
We calculated the self-energy including the off-diagonal terms for the 24 bands around the Fermi level. 
For the entangled bands, we employed the band disentanglement procedure of Ref.~\onlinecite{Miyake09}.
}

\section{\rr{Interstitial $s$ orbital near the Fermi level}}
\label{sec:interstital}

\begin{figure}[tbhp]
\vspace{0cm}
\begin{center}
\includegraphics[width=0.45\textwidth]{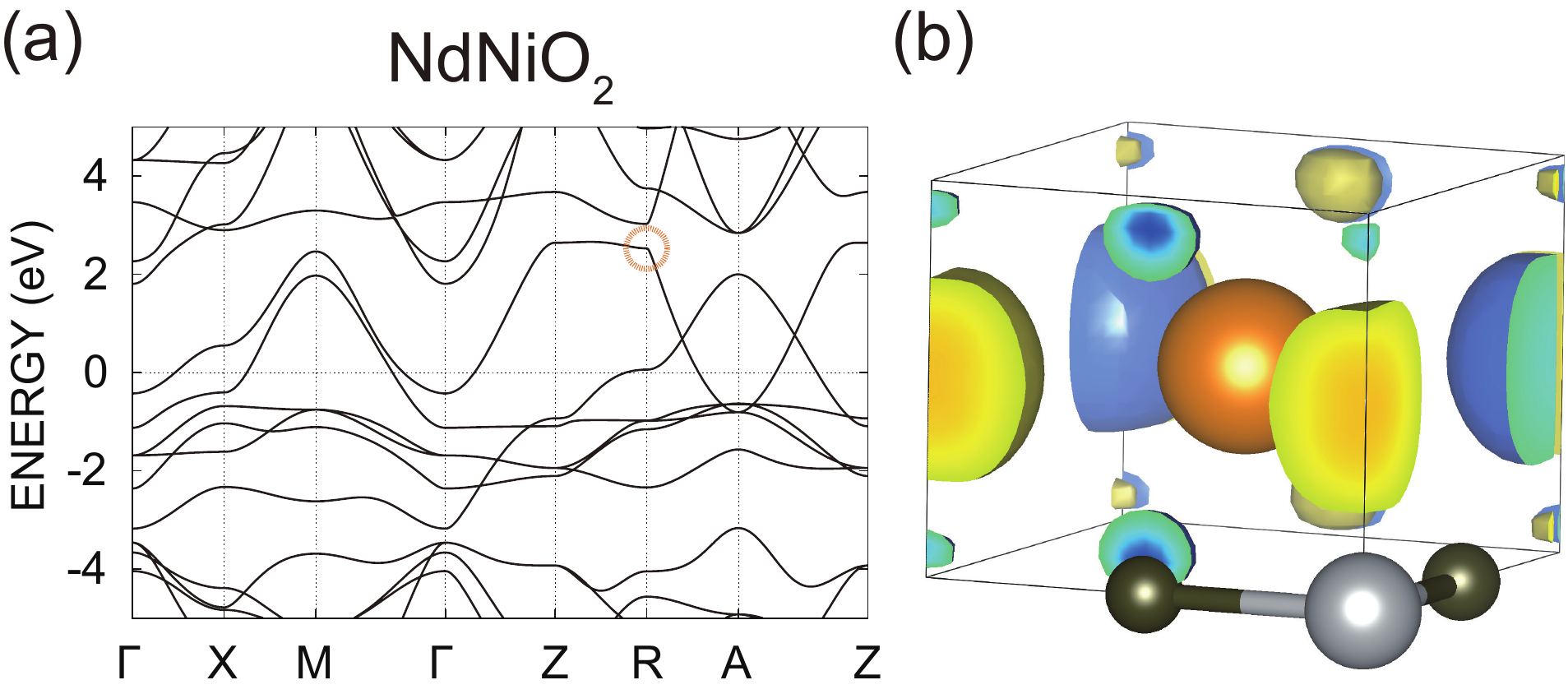}
\caption{
\rr{
(a) Electronic band structure of NdNiO$_2$.
The energy is measured from the Fermi level.
(b) Bloch function of the interstitial $s$ band at the R point.
The result shown in (b) corresponds to the state indicated by the orange open circle in (a).
}
}
\label{NdNiO2_Rpoint}
\end{center}
\end{figure}

\rr{
Figure~\ref{NdNiO2_Rpoint}(a) shows the electronic band structure of NdNiO$_2$.
It is characteristic that the interstitial $s$ band exists near the Fermi level (see Fig.\ref{NdNiO2_Rpoint}(b)).
The $s$ orbital is surrounded by the cation, which reduces the electric potential of the interstitial region.
}

\section{\rr{Electronic band structure in the GW approximation}}
\label{sec:GW}

\begin{figure}[tbhp]
\vspace{0cm}
\begin{center}
\includegraphics[width=0.4\textwidth]{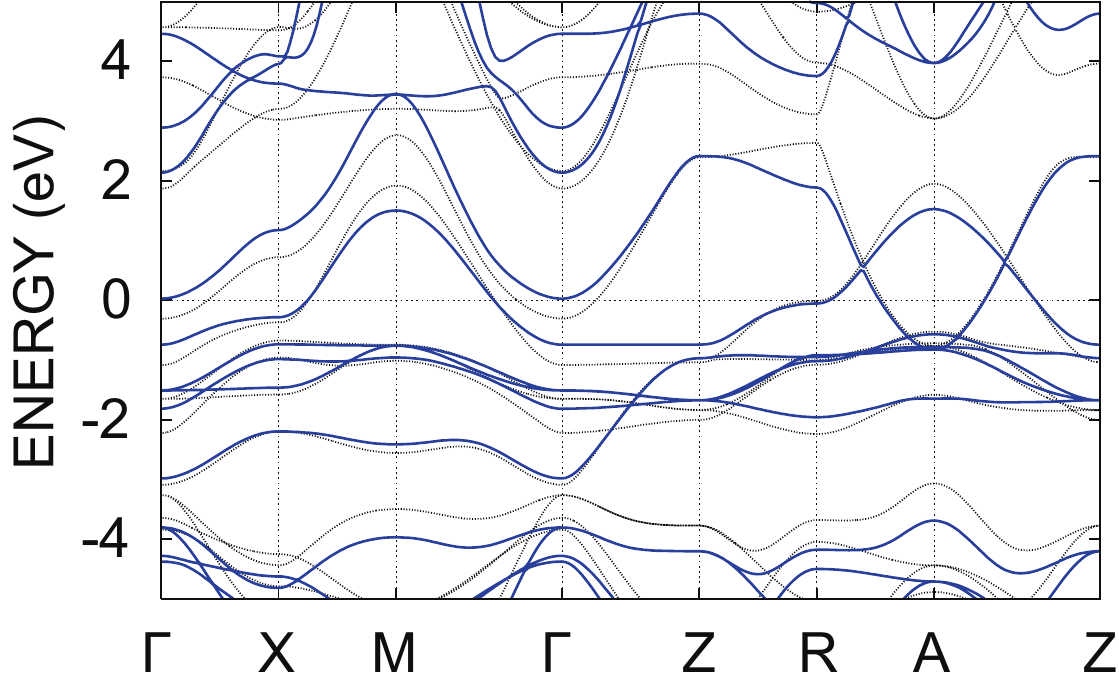}
\caption{
\rr{
Electronic band structures of LaNiO$_2$ in the GWA (blue solid line) and the GGA-DFT (black dotted line).
The energy is measured from the Fermi level.
}
}
\label{LaNiO2_GW_band}
\end{center}
\end{figure}

\rr{
In the GGA calculation of NdNiO$_2$, a Fermi pocket originating from the Nd $5d_{3z^2-r^2}$ orbital exists at the $\Gamma$ point. 
The bottom of the Nd $5d_{3z^2-r^2}$ band is close to the Fermi level (at $-400$ meV).
Given that the GGA tends to underestimate the band gap, such a small Fermi pocket is expected to disappear by a more accurate treatment of electron correlations.
We have confirmed this point by performing a calculation based on the GWA,
whose computational details are described in Appendix \ref{Appendix_GWA}.
Figure~\ref{LaNiO2_GW_band} shows the electronic band structure of LaNiO$_2$ in the GWA.
We see that the extra Fermi pocket originating from the $5d_{3z^2-r^2}$ orbital at the $\Gamma$ point indeed disappears.
By contrast, the size of the Fermi pocket around the A point is almost the same as that in the GGA calculation.
}

\section{Structural parameters and X-ray diffraction patterns of layered nickelates}
\label{sec:structure_and_xray}
\begin{figure}[htbp]
\centering
\includegraphics[width=0.45\textwidth]{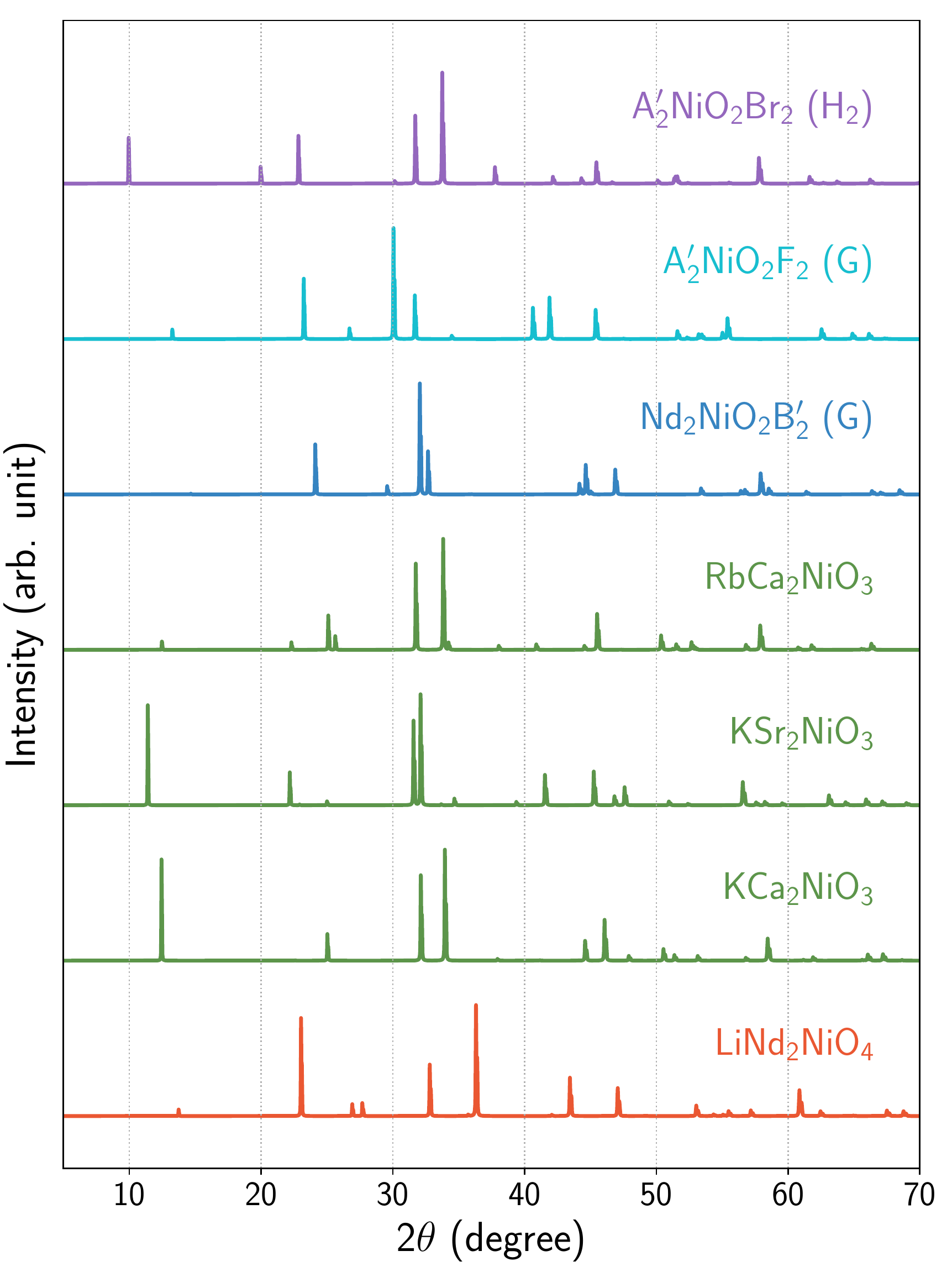}
\caption{X-ray powder diffraction patterns of several dynamically stable nickelates calculated by using the optimized structures. Different colors are used for different structural prototypes.}
\label{fig:xray}
\end{figure}

The complete list of the structural parameters obtained after the structural optimization based on PBEsol is given in Tables \ref{table:structures_1214_PBEsol}--\ref{table:structures_H2type}. In these tables, the dynamically stable compositions are shown in boldface. For some dynamically stable nickelates, the powder X-ray diffraction patterns are calculated by using \textsc{VESTA}~\cite{Momma_2011} and shown in Fig.~\ref{fig:xray}. X-ray diffraction patterns of the other nickelates can be calculated easily from the structure data shown in the tables.

\section{Structural parameters and phonons of layered palladates}
\label{sec:structure_and_phonon_Pd}
\begin{figure*}[hp]
\centering
\includegraphics[width=0.99\textwidth]{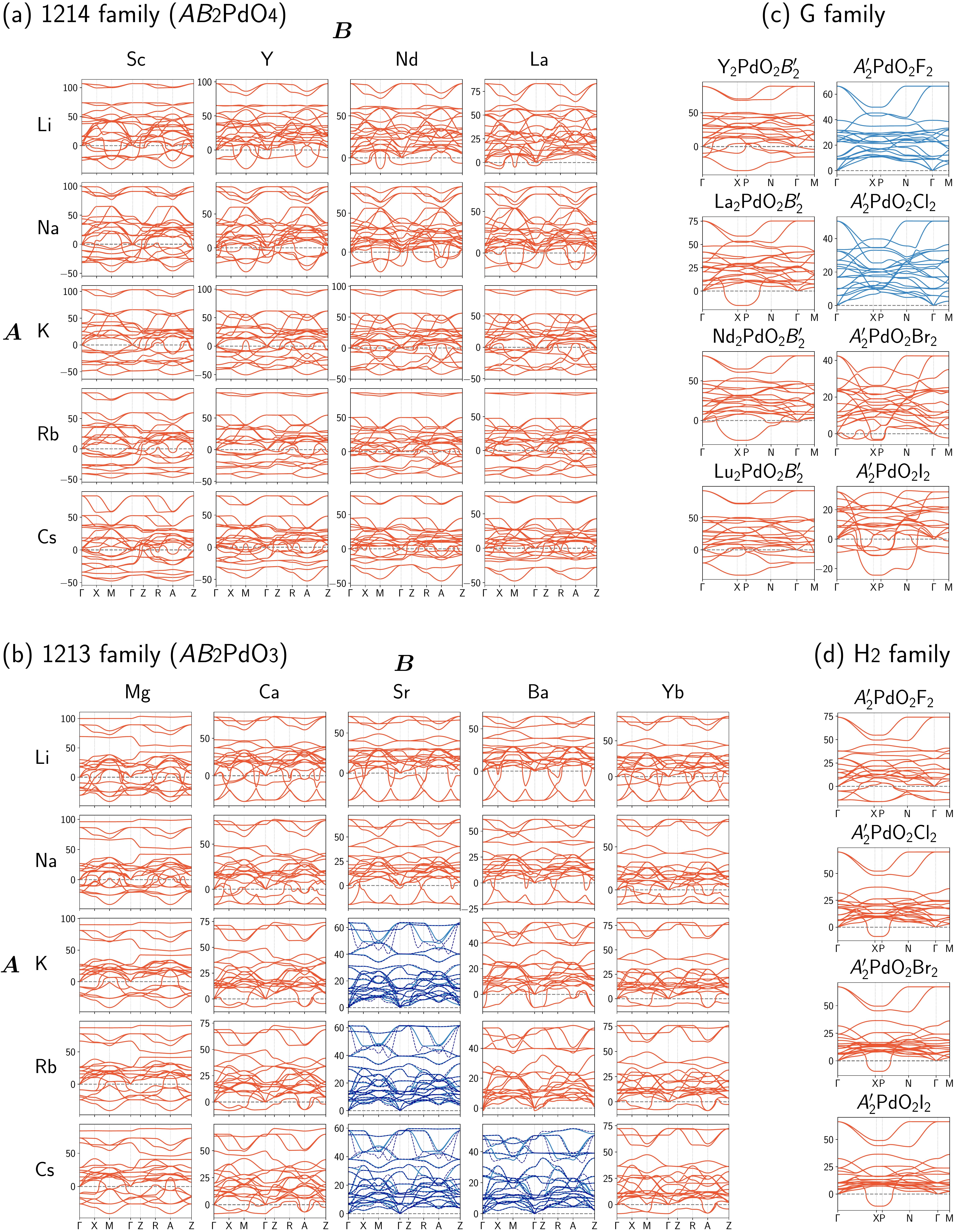}
\caption{Phonon dispersion curves of (a) 1214-family, (b) 1213-family, (c) G-family, and (d) H$_{2}$-family palladates calculated based on the PBEsol functional. 
We take $A'$ and $B'$ atoms to be 
Ba$_{0.5}$La$_{0.5}$ and O$_{0.5}$F$_{0.5}$, respectively.
The unit of the vertical axes is meV. For the 1214 and 1213 families, results based on the 2$\times$2$\times$2 and 4$\times$4$\times$2 supercells are shown with solid and dashed lines, respectively. Different colors are used for dynamically stable (blue) and unstable (red) compositions.}
\label{fig:Palladate_phonon}
\end{figure*}

The structure optimization and phonon calculation of the 57 studied palladates were performed by using exactly the same computational conditions as nickelates described in Appendix \ref{subsec_structure_opt} and \ref{subsec_phonon}.
The structural parameters of the studied palladates obtained after the structural optimization based on PBEsol are given in Tables \ref{table:structures_1214_PBEsol_Pd}--\ref{table:structures_H2type_Pd}. In these tables, the dynamically stable compositions are shown in boldface. Also, the calculated phonon dispersion curves of the 57 palladates are summarized in Fig.~\ref{fig:Palladate_phonon}.

\begin{table}[tbh]
\caption{Structural parameters of the \ptii structures ($P4/mmm$) optimized with PBEsol. The fractional coordinates of each atom are as follows: $A$:(0,0,$\frac{1}{2}$), $B$($\frac{1}{2}$,$\frac{1}{2}$,$\pm z$), Ni:(0,0,0), O:(0,$\frac{1}{2}$,0), ($\frac{1}{2}$,0,0), ($\frac{1}{2}$,0,$\frac{1}{2}$), (0,$\frac{1}{2}$,$\frac{1}{2}$). Dynamically stable compositions are shown in bold.}
\label{table:structures_1214_PBEsol}
\begin{ruledtabular}
\begin{tabular}{rcccc}
Composition & $a$ (\AA) & $c$ (\AA) & $c/a$ & $z$ \\
\hline
LiSc$_{2}$NiO$_{4}$ &  3.6725 & 5.6888 & 1.5490 & 0.2628 \\
LiY$_{2}$NiO$_{4}$ &  3.7799 & 6.1254 & 1.6205 & 0.2596 \\
\textbf{LiNd$_{2}$NiO$_{4}$} &  3.8584 & 6.4398 & 1.6691 & 0.2587 \\
\textbf{LiLa$_{2}$NiO$_{4}$} &  3.9035 & 6.5826 & 1.6863 & 0.2584 \\
NaSc$_{2}$NiO$_{4}$ &  3.8367 & 5.6984 & 1.4852 & 0.2614 \\
NaY$_{2}$NiO$_{4}$ &  3.9325 & 6.1000 & 1.5512 & 0.2580 \\
NaNd$_{2}$NiO$_{4}$ &  4.0061 & 6.3741 & 1.5911 & 0.2572 \\
NaLa$_{2}$NiO$_{4}$ &  4.0508 & 6.4935 & 1.6030 & 0.2569 \\
KSc$_{2}$NiO$_{4}$ &  4.0863 & 5.6913 & 1.3928 & 0.2515 \\
KY$_{2}$NiO$_{4}$ &  4.1774 & 6.0825 & 1.4561 & 0.2530 \\
KNd$_{2}$NiO$_{4}$ &  4.2398 & 6.3125 & 1.4889 & 0.2529 \\
KLa$_{2}$NiO$_{4}$ &  4.2813 & 6.4240 & 1.5005 & 0.2538 \\
RbSc$_{2}$NiO$_{4}$ &  4.1803 & 5.6927 & 1.3618 & 0.2399 \\
RbY$_{2}$NiO$_{4}$ &  4.2806 & 6.0507 & 1.4135 & 0.2449 \\
RbNd$_{2}$NiO$_{4}$ &  4.3402 & 6.2683 & 1.4442 & 0.2458 \\
RbLa$_{2}$NiO$_{4}$ &  4.3769 & 6.4123 & 1.4650 & 0.2458 \\
CsSc$_{2}$NiO$_{4}$ &  4.3039 & 5.7331 & 1.3321 & 0.2305 \\
CsY$_{2}$NiO$_{4}$ &  4.3887 & 6.0389 & 1.3760 & 0.2369 \\
CsNd$_{2}$NiO$_{4}$ &  4.4333 & 6.2465 & 1.4090 & 0.2376 \\
CsLa$_{2}$NiO$_{4}$ &  4.4588 & 6.4108 & 1.4378 & 0.2366 \\
\end{tabular}
\end{ruledtabular}
\end{table}

\begin{table}[tbh]
\caption{Structural parameters of the \pti structures ($P4/mmm$) optimized with PBEsol. The fractional coordinates of each atom are as follows: $A$:(0,0,$\frac{1}{2}$), $B$($\frac{1}{2}$,$\frac{1}{2}$,$\pm z$), Ni:(0,0,0), O:(0,$\frac{1}{2}$,0), ($\frac{1}{2}$,0,0), ($\frac{1}{2}$,$\frac{1}{2}$,$\frac{1}{2}$). Dynamically stable compositions are shown in bold.}
\label{table:structures_1213_PBEsol}
\begin{ruledtabular}
\begin{tabular}{rcccc}
Composition & $a$ (\AA) & $c$ (\AA) & $c/a$ & $z$ \\
\hline
LiMg$_{2}$NiO$_{3}$ &  3.7236 & 6.2398 & 1.6757 & 0.1997 \\
LiCa$_{2}$NiO$_{3}$ &  3.8164 & 7.1632 & 1.8770 & 0.1982 \\
LiSr$_{2}$NiO$_{3}$ &  3.8740 & 7.8523 & 2.0269 & 0.2006 \\
LiBa$_{2}$NiO$_{3}$ &  3.8923 & 8.6489 & 2.2221 & 0.2054 \\
LiYb$_{2}$NiO$_{3}$ &  3.7824 & 6.8848 & 1.8202 & 0.1906 \\
NaMg$_{2}$NiO$_{3}$ &  3.7621 & 6.2659 & 1.6655 & 0.1989 \\
NaCa$_{2}$NiO$_{3}$ &  3.8580 & 7.1527 & 1.8540 & 0.1966 \\
NaSr$_{2}$NiO$_{3}$ &  3.9196 & 7.8278 & 1.9971 & 0.1987 \\
NaBa$_{2}$NiO$_{3}$ &  3.9428 & 8.6133 & 2.1846 & 0.2033 \\
NaYb$_{2}$NiO$_{3}$ &  3.8210 & 6.8729 & 1.7987 & 0.1892 \\
KMg$_{2}$NiO$_{3}$ &  3.8379 & 6.2828 & 1.6371 & 0.1971 \\
\bf KCa$_{2}$NiO$_{3}$ &  3.9372 & 7.1090 & 1.8056 & 0.1931 \\
\bf KSr$_{2}$NiO$_{3}$ &  4.0047 & 7.7557 & 1.9366 & 0.1948 \\
\bf KBa$_{2}$NiO$_{3}$ &  4.0375 & 8.5135 & 2.1086 & 0.1997 \\
\bf KYb$_{2}$NiO$_{3}$ &  3.8901 & 6.8390 & 1.7581 & 0.1867 \\
RbMg$_{2}$NiO$_{3}$ &  3.8881 & 6.2946 & 1.6189 & 0.1967 \\
\bf RbCa$_{2}$NiO$_{3}$ &  3.9834 & 7.0911 & 1.7801 & 0.1922 \\
\bf RbSr$_{2}$NiO$_{3}$ &  4.0513 & 7.7240 & 1.9065 & 0.1937 \\
\bf RbBa$_{2}$NiO$_{3}$ &  4.0931 & 8.4596 & 2.0668 & 0.1982 \\
\bf RbYb$_{2}$NiO$_{3}$ &  3.9331 & 6.8240 & 1.7350 & 0.1860 \\
CsMg$_{2}$NiO$_{3}$ &  3.9650 & 6.3041 & 1.5899 & 0.1960 \\
CsCa$_{2}$NiO$_{3}$ &  4.0494 & 7.0750 & 1.7472 & 0.1910 \\
CsSr$_{2}$NiO$_{3}$ &  4.1140 & 7.6947 & 1.8704 & 0.1922 \\
CsBa$_{2}$NiO$_{3}$ &  4.1662 & 8.4015 & 2.0166 & 0.1962 \\
CsYb$_{2}$NiO$_{3}$ &  3.9943 & 6.8090 & 1.7047 & 0.1853 \\
\end{tabular}
\end{ruledtabular}
\end{table}

\begin{table}[tbh]
\caption{Structural parameters of the studied G-type $A_{2}$NiO$_{2}B'_{2}$ and $A'_{2}$NiO$_{2}B_{2}$ nickelates ($I4/mmm$) obtained with PBEsol functional. The $A$ ($A'$) atoms occupy the 4e site $(0,0,\pm z)$, and Ni atoms, O atoms in the NiO$_{2}$ layer, and $B'$ ($B$) atoms in the $AB'_{2}$ ($A'B_{2}$) layer are located at the 2a, 4c, and 4d sites, respectively. Dynamically stable compositions are shown in bold.}
\label{table:structures_Gtype1}
\begin{ruledtabular}
\begin{tabular}{rcccc}
Composition & $a$ (\AA) & $c$ (\AA) & $c/a$ & $z$ \\
\hline
Y$_{2}$NiO$_{2}B'_{2}$  &  3.7913 & 11.5016 & 3.0336 & 0.3615 \\
\bf La$_{2}$NiO$_{2}\bm{B}'_{2}$ &  3.9289 & 12.4049 & 3.1573 & 0.3637 \\
\bf Nd$_{2}$NiO$_{2}\bm{B}'_{2}$ &  3.8731 & 12.0737 & 3.1173 & 0.3627 \\ 
Lu$_{2}$NiO$_{2}B'_{2}$ &  3.7578 & 11.2515 & 2.9942 & 0.3613 \\ \\
\bf $\bm{A}'_{2}$NiO$_{2}$F$_{2}$  & 3.9928 & 13.3419 & 3.3415 & 0.3725 \\
$A'_{2}$NiO$_{2}$Cl$_{2}$ & 4.2286 & 14.8511 & 3.5121 & 0.3898 \\
$A'_{2}$NiO$_{2}$Br$_{2}$ & 4.3500 & 15.2682 & 3.5099 & 0.3944 \\
$A'_{2}$NiO$_{2}$I$_{2}$  & 4.4956 & 16.2578 & 3.6164 & 0.4042 \\
\end{tabular}
\end{ruledtabular}
\end{table}

\begin{table}[tbh]
\caption{Structural parameters of the studied H$_{2}$-type $A'_{2}$NiO$_{2}B_{2}$ nickelates ($I4/mmm$) obtained with PBEsol functional. The $A'$ and $B$ atoms occupy the 4e site $(0,0,\pm z)$, and Ni atoms and O atoms are located at the 2a and 4c sites, respectively. Dynamically stable compositions are shown in bold.}
\label{table:structures_H2type}
\begin{ruledtabular}
\begin{tabular}{rccccc}
Composition & $a$ (\AA) & $c$ (\AA) & $c/a$ & $z_{A'}$ & $z_B$ \\
\hline
$A'_{2}$NiO$_{2}$F$_{2}$  & 3.8709 & 14.1021 & 3.6431 & 0.3787 & 0.1998 \\
\bf $\bm{A}'_{2}$NiO$_{2}$Cl$_{2}$ & 3.9749 & 16.1162 & 4.0546 & 0.3953 & 0.1911 \\
\bf $\bm{A}'_{2}$NiO$_{2}$Br$_{2}$ & 3.9884 & 17.7762 & 4.4570 & 0.4053 & 0.1848 \\
\bf $\bm{A}'_{2}$NiO$_{2}$I$_{2}$  & 4.0113 & 20.8415 & 5.1957 & 0.4196 & 0.1735  \\
\end{tabular}
\end{ruledtabular}
\end{table}

\begin{table}[tbh]
\caption{Structural parameters of the 1214-family $AB_{2}$PdO$_{4}$ structures ($P4/mmm$) optimized with PBEsol. The fractional coordinates of each atom are as follows: $A$:(0,0,$\frac{1}{2}$), $B$($\frac{1}{2}$,$\frac{1}{2}$,$\pm z$), Pd:(0,0,0), O:(0,$\frac{1}{2}$,0), ($\frac{1}{2}$,0,0), ($\frac{1}{2}$,0,$\frac{1}{2}$), (0,$\frac{1}{2}$,$\frac{1}{2}$). All compositions are dynamically unstable.}
\label{table:structures_1214_PBEsol_Pd}
\begin{ruledtabular}
\begin{tabular}{rcccc}
Composition & $a$ (\AA) & $c$ (\AA) & $c/a$ & $z$ \\
\hline
LiSc$_{2}$PdO$_{4}$ &  3.8422 & 5.6742 & 1.4768 & 0.2669 \\
LiY$_{2}$PdO$_{4}$ &  3.9393 & 6.0890 & 1.5457 & 0.2616 \\
LiNd$_{2}$PdO$_{4}$ &  4.0064 & 6.3708 & 1.5902 & 0.2604 \\
LiLa$_{2}$PdO$_{4}$ &  4.0482 & 6.4982 & 1.6052 & 0.2598 \\
NaSc$_{2}$PdO$_{4}$ &  3.9669 & 5.7111 & 1.4397 & 0.2652 \\
NaY$_{2}$PdO$_{4}$ &  4.0567 & 6.0885 & 1.5009 & 0.2601 \\
NaNd$_{2}$PdO$_{4}$ &  4.1199 & 6.3370 & 1.5382 & 0.2589 \\
NaLa$_{2}$PdO$_{4}$ &  4.1603 & 6.4499 & 1.5503 & 0.2582 \\
KSc$_{2}$PdO$_{4}$ &  4.1886 & 5.7432 & 1.3711 & 0.2570 \\
KY$_{2}$PdO$_{4}$ &  4.2669 & 6.1080 & 1.4315 & 0.2560 \\
KNd$_{2}$PdO$_{4}$ &  4.3198 & 6.3158 & 1.4621 & 0.2554 \\
KLa$_{2}$PdO$_{4}$ &  4.3549 & 6.4244 & 1.4752 & 0.2561 \\
RbSc$_{2}$PdO$_{4}$ &  4.2834 & 5.7581 & 1.3443 & 0.2443 \\
RbY$_{2}$PdO$_{4}$ &  4.3665 & 6.0847 & 1.3935 & 0.2477 \\
RbNd$_{2}$PdO$_{4}$ &  4.4161 & 6.2783 & 1.4217 & 0.2477 \\
RbLa$_{2}$PdO$_{4}$ &  4.4491 & 6.4089 & 1.4405 & 0.2483 \\
CsSc$_{2}$PdO$_{4}$ &  4.3891 & 5.8343 & 1.3293 & 0.2310 \\
CsY$_{2}$PdO$_{4}$ &  4.4666 & 6.0926 & 1.3641 & 0.2384 \\
CsNd$_{2}$PdO$_{4}$ &  4.5077 & 6.2677 & 1.3904 & 0.2394 \\
CsLa$_{2}$PdO$_{4}$ &  4.5325 & 6.4134 & 1.4150 & 0.2386 \\
\end{tabular}
\end{ruledtabular}
\end{table}

\begin{table}[tbh]
\caption{Structural parameters of the 1213-family $AB_{2}$PdO$_{3}$ structures ($P4/mmm$) optimized with PBEsol. The fractional coordinates of each atom are as follows: $A$:(0,0,$\frac{1}{2}$), $B$($\frac{1}{2}$,$\frac{1}{2}$,$\pm z$), Pd:(0,0,0), O:(0,$\frac{1}{2}$,0), ($\frac{1}{2}$,0,0), ($\frac{1}{2}$,$\frac{1}{2}$,$\frac{1}{2}$). Dynamically stable compositions are shown in bold.}
\label{table:structures_1213_PBEsol_Pd}
\begin{ruledtabular}
\begin{tabular}{rcccc}
Composition & $a$ (\AA) & $c$ (\AA) & $c/a$ & $z$ \\
\hline
LiMg$_{2}$PdO$_{3}$ &  3.9993 & 6.2141 & 1.5538 & 0.2020 \\
LiCa$_{2}$PdO$_{3}$ &  4.0825 & 6.9881 & 1.7117 & 0.1973 \\
LiSr$_{2}$PdO$_{3}$ &  4.1412 & 7.6276 & 1.8419 & 0.1983 \\
LiBa$_{2}$PdO$_{3}$ &  4.1826 & 8.3593 & 1.9986 & 0.2021 \\
LiYb$_{2}$PdO$_{3}$ &  4.0450 & 6.7165 & 1.6604 & 0.1892 \\
NaMg$_{2}$PdO$_{3}$ &  4.0256 & 6.2507 & 1.5528 & 0.2017 \\
NaCa$_{2}$PdO$_{3}$ &  4.1102 & 6.9907 & 1.7008 & 0.1961 \\
NaSr$_{2}$PdO$_{3}$ &  4.1717 & 7.6125 & 1.8248 & 0.1968 \\
NaBa$_{2}$PdO$_{3}$ &  4.2202 & 8.3234 & 1.9723 & 0.2003 \\
NaYb$_{2}$PdO$_{3}$ &  4.0679 & 6.7281 & 1.6540 & 0.1886 \\
KMg$_{2}$PdO$_{3}$ &  4.0811 & 6.2876 & 1.5407 & 0.2004 \\
KCa$_{2}$PdO$_{3}$ &  4.1690 & 6.9745 & 1.6729 & 0.1932 \\
\bf KSr$_{2}$PdO$_{3}$ &  4.2324 & 7.5722 & 1.7891 & 0.1937 \\
KBa$_{2}$PdO$_{3}$ &  4.2901 & 8.2544 & 1.9241 & 0.1966 \\
KYb$_{2}$PdO$_{3}$ &  4.1168 & 6.7316 & 1.6352 & 0.1866 \\
RbMg$_{2}$PdO$_{3}$ &  4.1151 & 6.3123 & 1.5340 & 0.2001 \\
RbCa$_{2}$PdO$_{3}$ &  4.1987 & 6.9768 & 1.6616 & 0.1926 \\
\bf RbSr$_{2}$PdO$_{3}$ &  4.2611 & 7.5613 & 1.7745 & 0.1929 \\
RbBa$_{2}$PdO$_{3}$ &  4.3210 & 8.2297 & 1.9046 & 0.1958 \\
RbYb$_{2}$PdO$_{3}$ &  4.1437 & 6.7382 & 1.6261 & 0.1863 \\
CsMg$_{2}$PdO$_{3}$ &  4.1709 & 6.3377 & 1.5195 & 0.1997 \\
CsCa$_{2}$PdO$_{3}$ &  4.2439 & 6.9808 & 1.6449 & 0.1919 \\
\bf CsSr$_{2}$PdO$_{3}$ &  4.3015 & 7.5537 & 1.7561 & 0.1920 \\
\bf CsBa$_{2}$PdO$_{3}$ &  4.3609 & 8.2095 & 1.8825 & 0.1948 \\
CsYb$_{2}$PdO$_{3}$ &  4.1859 & 6.7456 & 1.6115 & 0.1859 \\
\end{tabular}
\end{ruledtabular}
\end{table}

\begin{table}[tbh]
\caption{Structural parameters of the studied G-type $A_{2}$NiO$_{2}B'_{2}$ and $A'_{2}$NiO$_{2}B_{2}$ palladates ($I4/mmm$) obtained with PBEsol functional. The $A$ ($A'$) atoms occupy the 4e site $(0,0,\pm z)$, and Pd atoms, O atoms in the PdO$_{2}$ layer, and $B'$ ($B$) atoms in the $AB'_{2}$ ($A'B_{2}$) layer are located at the 2a, 4c, and 4d sites, respectively. Dynamically stable compositions are shown in bold.}
\label{table:structures_Gtype1_Pd}
\begin{ruledtabular}
\begin{tabular}{rcccc}
Composition & $a$ (\AA) & $c$ (\AA) & $c/a$ & $z$ \\
\hline
Y$_{2}$PdO$_{2}B'_{2}$  &  3.9846 & 11.2469 & 2.8226 & 0.3572 \\
La$_{2}$PdO$_{2}B'_{2}$ &  4.1027 & 12.0643 & 2.9406 & 0.3610 \\
Nd$_{2}$PdO$_{2}B'_{2}$ &  4.0530 & 11.7608 & 2.9017 & 0.3593 \\ 
Lu$_{2}$PdO$_{2}B'_{2}$ &  3.9594 & 10.9881 & 2.7752 & 0.3568 \\ \\
\bf $\bm{A}'_{2}$PdO$_{2}$F$_{2}$  & 4.1914 & 12.8875 & 3.0747 & 0.3706 \\
\bf $\bm{A}'_{2}$PdO$_{2}$Cl$_{2}$ & 4.3623 & 14.6197 & 3.3514 & 0.3891 \\
$A'_{2}$PdO$_{2}$Br$_{2}$ & 4.4572 & 15.1064 & 3.3892 & 0.3938 \\
$A'_{2}$PdO$_{2}$I$_{2}$  & 4.6123 & 15.9378 & 3.4555 & 0.4020 \\
\end{tabular}
\end{ruledtabular}
\end{table}

\begin{table}[tbh]
\caption{Structural parameters of the studied H$_{2}$-type $A'_{2}$PdO$_{2}B_{2}$ palladates ($I4/mmm$) obtained with PBEsol functional. The $A'$ and $B$ atoms occupy the 4e site $(0,0,\pm z)$, and Ni atoms and O atoms are located at the 2a and 4c sites, respectively. All compositions are dynamically unstable.}
\label{table:structures_H2type_Pd}
\begin{ruledtabular}
\begin{tabular}{rccccc}
Composition & $a$ (\AA) & $c$ (\AA) & $c/a$ & $z_{A'}$ & $z_B$ \\
\hline
$A'_{2}$PdO$_{2}$F$_{2}$  & 4.1337 & 13.8978 & 3.3621 & 0.3807 & 0.2077 \\
$A'_{2}$PdO$_{2}$Cl$_{2}$ & 4.1773 & 15.6770 & 3.7529 & 0.3937 & 0.1946 \\
$A'_{2}$PdO$_{2}$Br$_{2}$ & 4.2058 & 16.5014 & 3.9235 & 0.3994 & 0.1919 \\
$A'_{2}$PdO$_{2}$I$_{2}$  & 4.2102 & 19.4025 & 4.6085 & 0.4148 & 0.1790 \\
\end{tabular}
\end{ruledtabular}
\end{table}

\section{Energy of the interstitial $s$ and the cation $d_{xy}$ orbitals}
\label{ene_table}

We summarize the energies of the interstitial $s$ orbital, the cation $d_{xy}$ orbital, and their bonding/antibonding states for the Ni-based compounds in Table~\ref{Ene_sd} and those for the Pd-based compounds in Table~\ref{Ene_Pd}.

\begin{table*}[ptb] 
\caption{
\red{
Energy of the interstitial $s$ orbital and the cation $d_{xy}$ orbital in the Ni-based compounds.
$E_{s}$ and $E_{d_{xy}}$ are the onsite potentials of the $s$ and $d_{xy}$ Wannier orbitals, respectively.
$E_b^k$ and $E_a^k$ are the energy levels of the bonding and antibonding states between the $s$ and $d_{xy}$ orbitals at the $k$ point, respectively.
$\Delta E_{sd}$ is the energy difference between the $s$ and $d_{xy}$ states, $\Delta E_{sd}=E_{s}-E_{d_{xy}}$.
$\bar{E}_{sd}$ is the average of $E_{s}$ and $E_{d_{xy}}$.
$\Delta E_{ba}^k$ is the energy difference between the bonding and antibonding bands at the $k$ point.
The unit for length is \AA.
The energy is measured with respect to the Fermi level in units of eV.}
} 
\renewcommand{\arraystretch}{1.2}
\begin{ruledtabular}
\begin{tabular}{c|cc|cccccc|cccc}
 & $a$  & $c$ & $E_s$ & $E_{d_{xy}}$ &   $E_b^A$ & $E_a^A$ & & & $\Delta E_{sd}$ &  $\bar{E}_{sd}$ & $\Delta E_{ba}^A$ &  \\ \hline  
LaNiO$_2$       & 3.959 & 3.375 & 4.024 & 3.080 & $-$1.584 & 9.625  & &  & 0.944 &3.552 & 11.209 & \\ 
NdNiO$_2$       & 3.921 & 3.281 & 4.017 & 3.046 & $-$1.567 & 9.270 & &  & 0.971 &3.532 & 10.837 & \\ 
\hline  
 & $a$  & $c$ & $E_s$ & $E_{d_{xy}}$ & $E_b^M$ & $E_a^M$ &  $E_b^A$ & $E_a^A$ & $\Delta E_{sd}$ & $\bar{E}_{sd}$ & $\Delta E_{ba}^M$ & $\Delta E_{ba}^A$  \\ \hline  
LiLa$_2$NiO$_4$ & 3.904 & 6.583 & 4.471 & 3.532 & $-$1.121 & 10.328 & 0.675 &  10.298 & 0.939 & 4.002 & 11.449 & 9.623 \\ 
LiNd$_2$NiO$_4$ & 3.858 & 6.440 & 4.514 & 3.435 & $-$1.141 & 9.832 & 0.724 & 9.957 & 1.079 & 3.975 & 10.973 & 9.233  \\ 
\hline  
KCa$_2$NiO$_3$  & 3.937 & 7.109 & 3.932 & 4.197 & 0.549 & 8.069 & $-$0.109 & 6.658 & $-$0.265 & 4.065 &7.520 & 6.767  \\ 
KSr$_2$NiO$_3$  & 4.005 & 7.756 & 4.388 & 4.787 & 0.247 & 10.270 & $-$0.272 & 10.796 & $-$0.399 &4.588 & 10.023 &11.068 \\
KBa$_2$NiO$_3$  & 4.037 & 8.514 & 4.825 & 4.108 & $-$0.345 & 11.407 &  $-$0.688 & 11.355 & 0.717 & 4.467 & 11.752 &  12.043  \\ 
KYb$_2$NiO$_3$  & 3.890 & 6.839 & 5.037 & 5.415 & 0.026 & 10.609 & $-$0.722 & 11.709 & $-$0.378 & 5.226 & 10.583 &12.431 \\ 
RbCa$_2$NiO$_3$ & 3.983 & 7.091 & 4.221 & 3.895 & 0.917 & 7.897 & 0.009 &  6.776 & 0.326 & 4.058 & 6.980 & 6.767 \\ 
RbSr$_2$NiO$_3$ & 4.051 & 7.724 & 4.307 & 4.740 &  0.559 & 10.043 & $-$0.191 &  11.066 & $-$0.433 & 4.524 & 9.484 & 11.257 \\ 
RbBa$_2$NiO$_3$ & 4.093 & 8.460 & 4.607 & 4.059 & $-$0.076 & 11.180 & $-$0.603 & 11.451 & 0.548 & 4.333 & 11.256 &12.054 \\ 
RbYb$_2$NiO$_3$ & 3.933 & 6.824 & 4.851 & 5.345 & 0.430 &  10.225 & $-$0.653 & 12.443 & $-$0.494 & 5.098 & 9.795 &13.096  \\
\hline  
 & $a$  & $c$ & $E_s$ & $E_{d_{xy}}$ & $E_b^X$ & $E_a^X$ &  $E_b^P$ & $E_a^P$ & $\Delta E_{sd}$ & $\bar{E}_{sd}$ & $\Delta E_{ba}^X$ & $\Delta E_{ba}^P$  \\ \hline 
La$_2$NiO$_2B'_2$ (G) & 3.929 & 12.405 & 4.688 & 3.143 &$-$0.121 & 9.768 & $-$0.114 & 9.772 & 1.545 & 3.916  &  9.889 & 9.886 \\
Nd$_2$NiO$_2B'_2$ (G) & 3.873 & 12.074 & 4.524 & 2.965 &$-$0.209 &  9.440 & $-$0.204 & 9.467& 1.559 & 3.745 & 9.649 & 9.671 \\ 
\hline  
$A'_2$NiO$_2$F$_2$ (G) & 3.993  & 13.342 & 4.785 & 3.795  & $-$0.301 & 11.312 & $-$0.295 & 11.381  & 0.990 & 4.290& 11.613 &11.676 \\
\hline  
$A'_2$NiO$_2$Cl$_2$ (H$_{2}$) & 3.975 & 16.116 & 5.399 & 3.504 & 1.002 & 8.754  & 0.979 & 8.671  & 1.895 &4.452 & 7.752 & 7.692 \\
$A'_2$NiO$_2$Br$_2$ (H$_{2}$) & 3.988 & 17.776 & 5.280 & 3.530 & 0.898 &  9.033 &0.874 & 9.004 & 1.750 & 4.405 & 8.135 & 8.130 \\ 
$A'_2$NiO$_2$I$_2$  (H$_{2}$) & 4.011 & 20.842 & 5.126 &3.698  & 0.973  & 9.472  & 0.960 & 9.269 & 1.428 &4.412  & 8.499 &8.309
\end{tabular}
\end{ruledtabular}
\label{Ene_sd} 
\end{table*}

\begin{table*}[ptb] 
\caption{
\red{
Energy of the interstitial $s$ orbital and the cation $d_{xy}$ orbital in the Pd-based compounds.
$E_{s}$ and $E_{d_{xy}}$ are the onsite potentials of the $s$ and $d_{xy}$ Wannier orbitals, respectively.
$E_b^k$ and $E_a^k$ are the energy levels of the bonding and antibonding states between the $s$ and $d_{xy}$ orbitals at the $k$ point, respectively.
$\Delta E_{sd}$ is the energy difference between the $s$ and $d_{xy}$ states, $\Delta E_{sd}=E_{s}-E_{d_{xy}}$.
$\bar{E}_{sd}$ is the average of $E_{s}$ and $E_{d_{xy}}$.
$\Delta E_{ba}^k$ is the energy difference between bonding and antibonding bands at the $k$ point.
The unit for length is \AA.
The energy is measured with respect to the Fermi level in units of eV.}
} 
\renewcommand{\arraystretch}{1.2}
\begin{ruledtabular}
\begin{tabular}{c|cc|cccccc|cccc}
 & $a$  & $c$ & $E_s$ & $E_{d_{xy}}$ & $E_b^M$ & $E_a^M$ &  $E_b^A$ & $E_a^A$ & $\Delta E_{sd}$ & $\bar{E}_{sd}$ & $\Delta E_{ba}^M$ & $\Delta E_{ba}^A$ \\ 
\hline  
KSr$_2$PdO$_3$  & 4.232 & 7.572  & 5.002 & 4.854 &  0.744  & 9.517 &  0.102 & 10.586 &0.148  & 4.928 & 8.773 & 10.484 \\ 
RbSr$_2$PdO$_3$ & 4.261 & 7.561  & 4.302 & 4.736 & 1.070 & 9.531 & 0.191 & 11.041 & $-$0.434 & 4.519 & 8.461 & 10.850 \\
CsSr$_2$PdO$_3$  & 4.302 & 7.554 & 4.359 & 4.667  &  1.565  &  9.139 &  0.312  &  11.114 & $-$0.308 &4.513  &7.574  &10.802   \\ 
CsBa$_2$PdO$_3$  &  4.361 & 8.210  & 4.178 & 4.266 & 0.746 &  9.427  & $-$0.196 & 10.910 & $-$0.088 & 4.222 & 8.681 & 11.106 \\ \hline
 & $a$  & $c$ & $E_s$ & $E_{d_{xy}}$ & $E_b^X$ & $E_a^X$ &  $E_b^P$ & $E_a^P$ & $\Delta E_{sd}$ & $\bar{E}_{sd}$ & $\Delta E_{ba}^X$ & $\Delta E_{ba}^P$  \\ 
\hline 
$A'_2$PdO$_2$F$_2$ (G) & 4.191 & 12.888 & 4.772  & 3.271 & $-$0.163  & 9.860 & $-$0.167 & 9.750 & 1.501 & 4.022 & 10.023 &9.917 \\
$A'_2$PdO$_2$Cl$_2$ (G) & 4.362 & 14.620  & 4.836 & 3.679 & 0.552 & 9.709 & 0.535 &  9.342 & 1.157 &4.258 & 9.157 & 8.807 
\end{tabular}
\end{ruledtabular}
\label{Ene_Pd} 
\end{table*}

\clearpage

\bibliography{apssamp}

\end{document}